# Wind-Driven Gas Networks and Star Formation in Galaxies: Reaction-Advection Hydrodynamic Simulations

David Chappell and John Scalo
*Astronomy Department, University of Texas, Austin, TX 78712*

**ABSTRACT**
The effects of wind-driven star formation feedback on the spatio-temporal organization of stars and gas in galaxies is studied using two-dimensional intermediate-representational quasi-hydrodynamical simulations. The model retains only a reduced subset of the physics, including mass and momentum conservation, fully nonlinear fluid advection, inelastic macroscopic interactions, threshold star formation, and momentum forcing by winds from young star clusters on the surrounding gas. Expanding shells of swept-up gas evolve through the action of fluid advection to form a "turbulent" network of interacting shell fragments whose overall appearance is a web of filaments (in two dimensions). A new star cluster is formed whenever the column density through a filament exceeds a critical threshold based on the gravitational instability criterion for an expanding shell, which then generates a new expanding shell after some time delay. A filament-finding algorithm is developed to locate the potential sites of new star formation.

The major result is the dominance of multiple interactions between advectively-distorted shells in controlling the gas and star morphology, gas velocity distribution and mass spectrum of high mass density peaks, and the global star formation history. The gas morphology strongly resembles the model envisioned by Norman & Silk (1980), and observations of gas in the LMC and in local molecular clouds. The dependence of the frequency distribution of present-to-past average global star formation rate on a number of parameters is investigated. Bursts of star formation only occur when the *time-averaged* star formation rate per unit area is low, or the system is small. Percolation does not play a role. The broad distribution observed in late-type galaxies can be understood as a result of either small size or small metallicity, resulting in larger shell column densities required for gravitational instability. The star formation rate increases with density, but dependences on gas velocity dispersion and average shell column density suggest that the dependence is multivariate. The distribution of gas velocities exhibits exponential tails over a broad range of parameter values and the velocity distribution for gas in filaments is nearly exponential. Decay simulations with no star formation suggest that the exponential tails are due to multiple shell interactions, not individual stellar winds. The cloud mass spectra, estimated using a simplified version of the structure tree method, tend to be power laws at the higher-mass end, with an index that is nearly independent of the star formation activity or model parameters. Kinetic energy decay in simulations without star formation yields a $t^{-1}$ dependence. We discuss how the simulations can be viewed in the context of various incomplete conceptual models, including collisional cloud coalescence, wind-driven turbulence, propagating star formation, forced mass-conserving Burgers turbulence, and granular fluids.

**Key words:** hydrodynamics, turbulence, stars: formation, ISM: bubbles, galaxies: ISM

## 1 INTRODUCTION

The collective behavior of star formation in disk galaxies appears to involve interactions between many stellar and interstellar processes acting over a large range of temporal and spatial scales. Central to understanding the star formation process is the question of how spatial and velocity structure is generated in the interstellar gas through the action of instabilities, magnetic fields, turbulence, star-formation, shocks, etc., how that structure leads to the creation of dense star-forming gas clouds, and how the resulting young stars act as feedback to the gas structure. The present paper attempts to address this question through the use of an "intermediate-level representation" of the hydrodynamics, in which gas structure and star formation are controlled by the interactions of wind-driven shells subject to nonlinear advection and a prescription for star formation in shells. These interactions lead to a "turbulent" evolving network or web of filaments (in two dimensions).

There is now considerable observational evidence supporting the view that some form of "turbulence" is an important organizing process at work in the ISM over a wide range of spatial scales, although the physical nature and source of this "turbulence" remain unclear. Molecular line widths are usually much broader than the predicted thermal widths, indicating that the gas motions are supersonic (e.g. Larson 1981). Furthermore, the line widths scale with the size of the region as a power-law (Larson 1981; see Elmegreen & Falgarone 1996 for more current results), reminiscent of incompressible turbulence or a field of discontinuities, although striking exceptions exist (e.g. Plume et al. 1997). Evidence for a turbulent ISM is found at very small scales in pulsar observations (Armstrong et al. 1995), and turbulence is often argued to provide a source of pressure in supporting the galactic disk (e.g. Ruzmaikin et al. 1988, McKee 1990, Lockman & Gehman 1991, and Falgarone et al. 1992). The observed gas morphology is also suggestive of a dynamic process at work. Molecular line, IRAS, HI, and extinction maps reveal irregular, often filamentary, scale-free gas distributions on scales from 0.01 pc to hundreds of parsecs (e.g. Scalo 1985, Falgarone & Phillips 1991, Elmegreen & Falgarone 1996, Chappell & Scalo 2000). This complex appearance suggests that the gas is in a highly agitated state and has not relaxed into equilibrium structures. A common interpretation of this observation is that the gas is "turbulent."

A likely organizing process at work in the ISM is the action of high-mass stars and protostellar winds on the surrounding gas and on future generations of star formation. Winds and supernovae lead to the formation of shock waves which expand into the ambient gas. If the shocked gas has time to cool, a compressed gas shell forms behind the shock, which sweeps up the interstellar gas as it expands outward. Such shells may drive gas motions and turbulence in the ISM if their kinetic energy and momentum is efficiently transferred to the surrounding gas. For example, Ruzmaikin et al. (1988) showed that the power input from supernovae may be sufficient to account for the overall interstellar cloud velocity dispersion. Abbot (1982) and Castor (1993) estimated and compared the energy input from various massive star sources in the Milky Way, and Tarrab (1983) did the same for nearby galaxies. Norman & Ferrara (1997) calculated in more detail the spectrum of energy input from H II regions, supernovae, and superbubbles. This idea is the basis of several self-regulating star formation scenarios. Norman & Silk (1980) suggested that, for low-mass star forming regions, this energy could maintain turbulence in the parent cloud, preventing an overall gravitational collapse and limiting star formation. They viewed the turbulence as consisting of a network of old shell fragments described by a cloud fluid and argued that star formation may persist if the shell fragments coalesce to form more massive gravitationally unstable clouds, similar in spirit to the larger-scale simulation models to be presented here. Observational support for this view is found in regions of vigorous star formation, in which the gas is often found to be organized into complex, possibly turbulent, networks of H II region-driven, wind-blown, or supernova shells (e.g. Braunsfurth & Feitzinger 1983, Meaburn 1984, Bruhweiler et al. 1991, Chu & Kennicutt 1994, and Kim et al. 1998 for studies of the the Large Magellanic Cloud; Chini et al. 1997, Bally et al. 1999 for studies of the Orion and Circinus clouds, respectively).

Expanding gas shells may also mediate propagating star formation. According to Elmegreen and Lada (1977), the shells may become gravitationaly unstable and lead to the formation of a new stellar generation. (See Vishniac 1994, Elmegreen 1994, 1999, and references therein for a more detailed discussion of the problem.) The winds and supernovae from the new stars propel the process such that star formation may propagate through a cloud or galaxy. Their model was originally developed to describe the spread of high-mass star formation through a molecular cloud. Propagating star formation theories have since been developed for shells driven by protostellar winds from low-mass stars (Norman & Silk 1980) and for large-scale propagating star formation mediated by superbubbles created around entire OB clusters (e.g. Elmegreen 1994, Jungwiert & Palous 1994, Silich et al. 1996, Ehlerova et al. 1997). It has also been suggested that star formation could proceed at the intersections of converging flows in the turbulence as suggested by Elmegreen (1993) in a slightly different context. These scenarios suggest a *positive* feedback mechanism in which star formation leads to the birth of more stars. Observational evidence has been found for propagating star formation associated with superbubbles (Oey & Massey 1995, Lortet and Testor 1988, and Dopita et al. 1985), shells created by individual supernovae (Elmegreen 1989), and the compression of clouds on small scales by strong UV radiation fields from massive stars (Sugitani et al. 1995). Comprehensive surveys of the evidence are given by Elmegreen (1992, 1999).

These observations suggest a picture of the ISM in which shells of swept-up (i.e. compressed) ambient gas created around sites of vigorous star formation interact to form a turbulent background of old shell remnants. This picture has observational support in regions of active star formation. For example, in the giant HII region 30 Dor in the Large Magellanic Cloud, Chu & Kennicutt (1994) describe the morphology as a "complex network of expanding systems" and cataloged many shells over a broad range of spatial scales and velocities. The H I aperture synthesis map of the entire LMC by Kim et al. (1998) is dominated by filaments, shells, and holes. The detailed study of the smaller-scale Circinus molecular cloud by Bally et al. (1999; see also Dobashi, Sato, & Mizuno 1998) illustrates how the same type of dynamical morphology occurs due to "churning" by outflows from lower-mass protostars. In addition, shell interactions may be the primary mechanism controlling star formation in these regions. For example, Bruhweiler et al. (1991), and Chu & Kennicutt (1994) report evidence that star formation occurs at the intersections of supershells in 30 Dor. Furthermore, Hyland et al. (1992) suggest that star formation is maintained in this region because of shell interactions.

The recent HST/NICMOS study of 30 Dor in the LMC by Walborn et al. (1999), in conjuction with the morphological evidence in Chu & Kennicutt (1994) and Kim et al. (1998), strongly suggests that shell networks and propagating star formation are intimately related. This relation is one focus of the present work. This scenario of star formation mediated by a network of shells was first proposed by Norman & Silk (1980) for low-mass star-forming regions, and the observational results of Bally et al. (1999) for the Circinus molecular cloud strongly resembles this picture. While Norman & Silk (1980) incorporated star formation feedback processes, they modeled the turbulence through a one-zone cloud coagulation model which neglects all non-linear spatial hydrodynamic interactions and assumes equilibrium solutions which rely on balancing source and sink terms in a set of ordinary differential equations. Such "one-zone" models are unable to investigate either spatial or dynamic phenomena. The inclusion of nonlinearities, time delays, or multiple interacting "phases" into one-zone models frequently results in nonequilibrium behavior such as oscillations, bursts, or chaos. Nonequilibrium models have been applied to cloud fluid systems (Scalo & Struck-Marcell 1987, Vazquez & Scalo 1989), propagating star formation modeled as diffusion (Shore 1981, Ferrini & Marchesoni 1984, Korchagin et al. 1988), supernova-driven multicomponent ISM models (Ikeuchi & Tomita 1983, Ikeuchi et al. 1984), and self-regulating star formation systems (Parravano 1989). An important question regarding these models is the extent to which the resulting nonequilibrium behavior depends on the lack of spatial degrees of freedom in the model equations. More recently, Norman & Ferrara (1996) demonstrated the importance of star formation feedback by calculating the source spectrum of forcing due to HII regions, supernovae, and superbubbles. However, in the treatment of spatial coupling, they modeled transfer between scales in Fourier space by the advection term (we use "advection term" to refer to the $\mathbf{u} \cdot \nabla \mathbf{u}$ operator in the fluid momentum conservation equation) through a transfer function applicable for *incompressible* flows, an assumption which contradicts the observed supersonic motions in the ISM. Furthermore, the spatial distribution was only treated at the level of the two-point correlation function.

The main goal of the present work is to use simulations to study the spatial behavior caused by highly compressible advection in the presence of threshold star formation and stellar wind-driven expanding shells, and to investigate the global behavior that results from "opening up" the spatial degrees of freedom. We want to understand how stellar outflows "churn" (Bally et al. 1999) the ISM.

Several past studies have attempted to investigate the interaction between star formation and gas structure using hydrodynamic simulations. Chiang & Prendergast (1985) introduced a simulation in which stars are also modeled as a fluid, with the star formation rate proportional to the gas density to some power. In their simulations, a network of gas filaments surrounding regions of active star formation emerges as a consequence of a star-gas instability. Bregman et al. (1993) considered a similar model but also investigated a simulation oriented vertically in the galactic plane with the inclusion of a galactic gravitational potential. Vazquez-Semadeni et al. (1994) and Passot et al. (1995) studied a turbulent model which includes the effects of self-gravity, magnetic fields, rotation, stellar heating, and cooling and which models star formation as a threshold function of the gas density. Navarro & White (1993) also adopted a threshold star formation law in galaxy formation models. They found that their results depended sensitively on the prescription chosen for stellar energy injection.

While full hydrodynamic approaches provide the most detailed information on the star-gas interaction, large simulations including many physical processes require supercomputers and even then only a modest number of simulations can be run, and for a relatively short time. In complex problems that contain many underdetermined parameters, large regions of parameter space must be left unexplored. Furthermore, even if we were confident in the selection of parameter values and we somehow knew that the hydrodynamic simulations accurately reproduced the true star-gas dynamics, we would still not be guaranteed a satisfactory *theory* of star formation in the sense of an explanatory model. As Kaneko (1991) puts it "If one succeeds in reproducing the phenomena from the model equation, then what can one learn? One dangerous trap in computational physics is that one may be still in the same level as the direct observation of the complex phenomena itself. What one might obtain is just that the equations are correct and reasonable, without any *understanding* of the complex phenomena." In astrophysics the situation is a bit different since we do not have the luxury of directly observing the "phenomena" over time. Thus, simply being able to watch the simulated evolution of the ISM, for example, could in itself be very useful. However, numerous examples from fields including non-linear dynamics, turbulence, biology, ecology, and economics, have found that in studying complex systems which exhibit self-organization or the emergence of non-trivial correlations, the most detailed simulations do not always provide the greatest insight into the phenomena. These studies find that experimentation with the *level of representation* of a simulation can lead to new insight into the nature of the underlying dynamics and can play an important role in formulating new theories. A recurrent finding of these investigations is the result that the phenomenology, statistical behavior, scaling laws, etc., of many "complex" systems can be reproduced by simple models which possess the symmetries, conservation laws, nonlinearities, or spatial connectivities of the physical system without reference to the detailed form of the system.

One of the more successful examples of this approach was used by Yanagita and Kaneko (1995) to study Rayleigh-Bernard convection. They combined a simple donor-cell advection scheme with "rules" to simulate the effects of buoyancy and incompressibility and reproduced nearly all the phenomena observed in Rayleigh-Bernard convection experiments including the formation of convection rolls, oscillations, routes to chaos, and spatio-temporal intermittency. In fact, since their method is computationally efficient, they were able to explore the parameter space of the model, and discovered a new class of phenomena. The present paper is similar to this convection study with respect to the level of representation.

A number of papers have tried to model the spatial as-pect of the galactic star formation problem by circumventing the hydrodynamics, particularly in the area of propagating star formation. Perhaps the best-known is the stochastic self-propagating star formation (SSPSF) model (see Seiden & Gerola 1982 for a review) in which active star-forming sites spread stochastically over a spatial lattice. While this model abstracts the detailed physics of the propagation mechanism to a simple stochastic rule, it proposes a picture of the organization mechanism through the theoretical framework of percolation theory. Gerola and Seiden used this model to make predictions about the coherence of star formation in disk galaxies, in addition to global star formation histories in dwarf galaxies. Their model has been extended to include anisotropic effects (e.g. Jungwiert & Palous 1994) and 3-D geometries (Comins 1983). Lacking from the SSPSF picture is the potentially important role that the gas dynamics (in particular the nonlinearity of the advection operator and consequences of conservation laws) may have on the stellar organization. Also lacking from such models is an explicit treatment of the physics of the propagation process, which is simply imposed as a stochastic "rule." More ambitious cellular automaton-type models have been presented by Perdang & Lejeune (1996) and Lejeune & Perdang (1996), who have generalized the Gerola and Seiden approach to include a highly deformable grid in the presence of a prescribed velocity field with both rotational and non-rotational (diffusive) components, as well as several stellar evolutionary states. Particularly interesting is the ability of these models to produce spatial structures with fractal dimensions similar to those observed in the interstellar medium. However these models are still non-hydrodynamic and do not allow for any feedback between star formation and the (prescribed) velocity field. An even more detailed cellular automaton model for propagating star formation was presented by Gardiner et al. (1998), who include an N-body calculation of the gravitational interactions and star formation feedback. However the simulations are still non-hydrodynamic, and so the important effects of nonlinear fluid advection are omitted. Another recent non-hydrodynamic cellular automaton model, concentrating on radiative transfer coupling between cells, was presented by Rousseau, Chate, & Le Bourlot (1998).

Reaction-diffusion models of propagating star formation have also been studied. Shore (1983) introduced a model similar to the Lotka-Volterra equation in which the propagation of star formation is assumed to be controlled by a diffusion process. Feitzinger (1984), Neukirch & Feitzinger (1988), Neukirch & Hesse (1993), Nozakura & Ikeuchi (1984, 1988), and Korchagin & Ryabtsev (1992) have used similar reaction-diffusion descriptions to study the problem of pattern formation in the ISM. Patterns formed in reaction-diffusion models are often referred to as "dissipative structures," following Nicolis & Prigogine (1977). The main drawback of these models is their neglect of the underlying gas dynamics, in particular the replacement of the nonlinear advective spatial operator with a linear diffusion operator for the gas. This severe assumption suppresses whole classes of phenomena (e.g. shocks) and may inappropriately introduce others (e.g. Turing pattern formation). For this reason we believe that recent re-introductions of this model (Smolin 1996, Freund 1997) should be regarded with caution.

In this paper, we develop a model for stellar "churning" of the ISM which is structurally simpler than a full hydrodynamic simulation, but which retains the fundamental mass and momentum conservation laws and the nonlinear hydrodynamical effects of fluid advection. However, pressure is neglected, magnetic fields are ignored, and the effects of self-gravity are abstracted as a threshold rule for star formation based on a linear gravitational instability criterion. Instead of attempting to build a "complete" model of the ISM, we have chosen to investigate this hybrid or "reduced" model to study the organization and evolution of star formation driven by the interactions of a system of turbulent wind-blown shells. While the structure of this model is in a sense simplistic, it includes enough physics to be of interest from several theoretical perspectives including theories of propagating star formation, turbulence, self-regulation, cloud coalescence, and pattern formation in general.

The model developed in this paper simulates a system of interacting shells driven by winds and supernovae from massive stars, although the general results should be applicable to the case of clouds whose internal structure and motions are driven by protostellar or protocluster winds. Stars are born from the shells when the gas column density through a shell exceeds a threshold value, and their winds generate new expanding shells, perpetuating the propagation of star formation. The shell dynamics are controlled by fluid advection and obey global mass and momentum conservation. The simplest set of fluid equations satisfying these requirements are the continuity equation and the pressureless momentum equation. By neglecting pressure gradients, collisions between gas structures become completely inelastic, causing the colliding shells to "stick." This limit corresponds to a vanishing adiabatic index $\gamma$ which has been estimated to be considerably less than unity in the interstellar medium, at least at low and moderate densities (see Vazquez-Semadeni et al. 1996 and Scalo et al. 1998). It can also be interpreted as assuming that ram pressure dominates thermal pressure everywhere, i.e. that the motions are everywhere highly supersonic. (However these two interpretations are not equivalent, as recently shown by Passot & Vazquez-Semadeni 1998.) By neglecting the pressure gradient term and by only incorporating self-gravity in terms of a threshold condition on star formation based on a gravitational instability condition, the computational speed increases considerably, allowing us to better explore parameter space and test variations of the model. In addition, the simplification of the model equations allows for easier interpretation and opens up possibilities for constructing analytic theories of the dynamics.

We refer to this model as a "reaction-advection" system since the gas motions are driven by young star clusters (the formation of which depends on the gas properties) and evolve through the action of nonlinear fluid advection. We suggest that, like reaction-diffusion models, it may be generalized to study generic spatio-temporal pattern formation, but would be applicable to systems in which the transport process is best described by fluid advection rather than diffusion. The model is, in effect, equivalent to forced Burgers turbulence, although our "forcing" is tied to the physics of star formation and stellar winds rather than simply assuming some arbitrary (usually Gaussian) stochastic forcing, and we explicitly solve the mass continuity equation.

In section 2, we present the fluid equations, star forma-tion criteria, and stellar forcing prescription, and discuss the computational methods employed. In section 3, the spatial distribution is presented for model runs which include star formation feedback and for decay runs with no stellar forcing. Section 4 discusses the global properties of the simulations, including the distributions of present to past average SFR ratios, the gas velocity, and the cloud mass spectra. An interesting aspect of the simulations is that they can be viewed in terms of several more "highly-reduced" models such as collisional cloud coalescence or propagating star formation. A discussion of this "multiperspectival" aspect is given in section 5, followed by a summary.

## 2 THE MODEL

The motions of the interstellar gas in our model are controlled by fluid advection and momentum forcing due to star formation:

$$\frac{\partial \rho}{\partial t} + \nabla \cdot (\rho \vec{v}) = 0 \quad (1)$$

$$\frac{\partial \rho \mathbf{v}}{\partial t} + \nabla \cdot (\rho \mathbf{v}\mathbf{v}) = \sum_{x'} \frac{\mathbf{x} - \mathbf{x}'}{|\mathbf{x} - \mathbf{x}'|} \frac{p_w N_*(\mathbf{x}', t)}{\tau_w} \quad (2)$$

where $\rho$ is the gas surface density, $p_w$ is the total momentum input per massive star, $N_*(\mathbf{x}', t)$ is the number of stars per unit area injecting momentum at position $x'$, and $\tau_w$ is the duration of the momentum injection ($10^7$ yr here). Star formation is assumed to occur according to a shell column density threshold criterion meant to represent shell gravitational instability, as explained in section 2.2 below. In practice we also experimented with models that include the source term in the continuity equation (1), accounting for the gas lost to star formation. However because the timescale for this gas depletion is so large compared to the phenomena of interest, and in order to calculate averages and other statistical quantitiesfrom a stationary distribution, we have omitted the mass depletion term for the calculations reported here.

The momentum source term in eq. 2 only symbolically represents the finite difference procedure followed in the simulations. The position in question is $\mathbf{x}$. The sum is over the eight nearest neighbor cells, at positions $\mathbf{x}'$. The unit vector ensures that the momentum is directed toward the position $\mathbf{x}$. If there is a cluster at $\mathbf{x}'$, then $N_*(x')$ is the number of newly-formed stars at that position, per unit area. The number of stars formed in that cluster is computed from the mass in the cluster using an adopted IMF. The cluster mass is computed from the mass in the simulation cell times an assumed constant star formation efficiency. The momentum input $p_w = m(x')v$ is calculated using a constant assumed wind velocity, 40 km s$^{-1}$, at a distance corresponding to a cell size (7.8 pc in the simulations reported here), and $m(x')$ is the fraction of the mass released by the cluster at $\mathbf{x}'$ that enters the cell at $\mathbf{x}$, assuming the morphology of the wind is circular in two dimensions. (We do not yet address the interesting question of the effect of collimated outflows rather than spherical winds.) The division by $\tau_w$ signifies that this mass and momentum are redistributed over the lifetime of the wind, $10^7$ yr. Momentum is only injected after a time $\tau_d$ since the onset of gravitational instability (see section 2.2 below). The motion of the cluster at position x$'$ is taken into account when calculating velocities. Star formation is in effect externally specified by selecting a criterion, generally a function of $\rho$ and $\vec{v}$, as discussed below in section 2.2.

The effects of pressure gradients were excluded from the model for both computational and conceptual reasons. The relative importance of pressure to advection in the ISM may be estimated through dimensional analysis as

$$\frac{\mu^{-1}\nabla P}{\vec{v} \cdot \nabla \vec{v}} \approx \frac{\gamma c^2 L^{-1}}{v^2 L^{-1}} \approx \frac{\gamma}{M^2} \quad (3)$$

where $c$ is the adiabatic sound speed, $L$ is the characteristic scale of the flow, and $M$ is the characteristic Mach number. Since, as discussed above, $\gamma$ may be small and $M$ is large in much of the ISM, the effects of pressure are frequently dominated by advection. Several observational studies have found that macroscopic "turbulent" gas motions in the ISM dominate thermal pressure, especially at large scales (e.g. Falgarone et al. 1992) and that gas motions in the interstellar medium are frequently supersonic with Mach numbers from a few to 100 (e.g. Larson 1981) so the neglect of thermal pressure may be a good approximation. The utility of the Burgers equation for studying highly supersonic astrophysical flows has been demonstrated by Kofman & Raga (1992 and references therein) in the context of stellar jet structure. Still, it should be remembered that, with constant pressure, the system has some unique properties: e.g. there are no sound (pressure) waves to carry information and all compressions develop into discontinuities due to the lack of pressure gradient forces. Passot & Vazquez-Semadeni (1998) have used a model equation to show that the zero-pressure behavior is not equivalent to the limit $\gamma \to 0$, although the major difference they find is for the lowest-density regions, which are not of primary interest in the present work.

In addition, the effects of pressure cannot be neglected near shock discontinuities. In the present simulations the widths of filamentary structures formed by shocks and their interactions are instead controlled by the (unphysical) numerical diffusion, which sets the filament widths initially to a few cell widths and causes them to gradually fatten with time (see sec. 2.1 and Appendix A). Such an effect could mimic the effects of real turbulence within the filaments (if it exists), but the scaling of this physical effect is unknown, and so our filament thicknesses cannot be considered realistic. On the other hand, the critical quantity for the star formation threshold that we adopt is the column density *through* a filament at a given location, and this quantity should not depend much on the filament thickness. The identification of filaments and the calculation of their local column densities is described in sec. 2.3 below.

### 2.1 Donor-cell advection

We choose the donor-cell scheme to model fluid advection on a discrete lattice. The main advantage from our perspective is its straightforward representation which may be easily adapted to include momentum injection by star formation and, in a future study, moving grids to simulate galactic differential rotation. The donor-cell method may allow higher dynamic ranges in the fluid density than, for example, SPH codes (Durisen et al. 1986). However, the use of adaptive smoothing lengths in modern SPH codes reduces this disparity (see Kang et al. 1994). Also, since the method is explicit, expensive matrix inversion techniques are not required as in implicit methods (van Leer 1977). Donor-cell methods have been used by Norman et al. (1980), Burkert & Hensler (1988), and Boss (1993) in simulations of gravitational cloud collapse and by Yanagita & Kaneko (1995) in their study of convection.

We use a modified donor-cell scheme in which a planar function is used to approximate the gas density at each lattice site (see Appendix A). While the donor-cell scheme is an Eulerian technique with a fixed grid, it is most conveniently defined by imagining that on each time step, the $i^{th}$ cell is transported a distance $u^i \Delta t$, where $u^i$ is the fluid velocity. Each cell donates the portion of the fluid variable which overlaps its downstream neighbor. The advection scheme was constructed to minimize the effects of numerical diffusion in regions of the flow that are "ballistic" (which accounts for a large fraction of the life of a fluid parcel). In such regions numerical diffusion may be approximated by the $\partial^4/\partial_x^4$ operator, while near discontinuities, it is better represented by the $\partial^2/\partial_x^2$ operator, as shown in App. A. The effect of this prescription is that strong discontinuities quickly diffuse as $\Delta l \propto \Delta t^{1/2}$ until the discontinuity is spread typically over a few lattice sites, but then diffuses more slowly as $\Delta l \propto \Delta t^{1/4}$ once it has been softened. The details of the method and its associated numerical diffusion are given in Appendix A. The advantage of this approach is that it is more computationally efficient than full second-order techniques, and is substantially less diffusive than purely first-order methods. Standard tests of hydrodynamic codes (e.g. the shock tube problem) cannot be used, since they involve pressure. However the code does accurately conserve mass and momentum, and we have checked the effect of resolution between $128^2$ and $512^2$, as discussed below. However we point out that we did not carry out a conventional resolution convergence test, in which the number of resolution elements is varied with all other input quantities held constant. Instead, by varying the domain size $L$, we are increasing the resolution, but are also changing characteristic scales, for example the ratio of the size corresponding to the peak in the energy spectrum of initial conditions to the size of the domain, or the total energy injection by the winds over the entire domain. Nevertheless, it will be seen below that the average values of all the statistical quantities of interest, including velocity probability distributions, are fairly stable over this factor of four change in the size of the system at fixed grid spacing, indicating that the results are not sensitive to the resolution defined in this way. Most of the results to be discussed are for $256^2$ runs.

### 2.2 Star formation

We assume that star formation is driven by gravitational instabilities in expanding gas shells (e.g. Elmegreen 1994, Comoron & Torra 1994, Whitworth et al. 1994, McCray & Kafatos 1987) or in compressed layers between colliding gas flows (e.g. Elmegreen 1993, Vishniac 1994, Whitworth et al. 1994). Since neither pressure nor gravity are explicitly included in the simulations, we rely on analytic results based on linear stability analysis of the hydrodynamic equations to determine when the gas is unstable to collapse. The adoption of such star formation laws based on linear stability analysis or more ad hoc schemes such as the Schmidt law must be used even in fully hydrodynamic simulations involving more physical processes (see references given below), since the tremendous range of scales involved in the star formation process makes it unfeasible to follow the entire collapse. However our star formation criterion is different from previous work because the criterion depends on the column density through individual structures measured along the plane of the simulation.

In general, we assume that star formation proceeds when the growth rate of the fastest-growing mode $\omega_{\text{fast}}$ (which depends on the geometry and overall dynamics of the local gas distribution) exceeds a critical growth rate $\omega_c$:

$$\omega_{\text{fast}} > \omega_c. \qquad (4)$$

In analytic models of star formation which consider the stability of expanding shells, the value of $\omega_c$ usually is set to the age of the shell with the idea being that at earlier times the instability would not have had enough time to develop (Elmegreen 1994, Comeron & Torra 1994). The existence of a critical growth rate also results if collapsing gas clouds can be disrupted by passing shells or clouds (Franco and Cox 1983). Franco and Cox proposed that star formation proceeds only when the gas free-fall time becomes less than a "reagitation time" due to the passage of shells. We adopt this interpretation of the critical growth rate and further set $\omega_c$ to a constant throughout the simulation. This is an admittedly large simplification, but we do not wish to introduce any more free parameters or spatially-dependent processes than necessary, so that the interpretation of the results will be as unambiguous as possible, in keeping with our goal of an intermediate-level representation.

The stability criterion used for the majority of the simulations is based on the stability of an expanding, accreting sheet and is derived in Appendix B. This instability condition depends on the velocity dispersion within the shocked gas layer. However, the structure of the the gas velocity field behind shock fronts is likely to be disordered and possibly turbulent since nonuniformities in the pre-shock gas distribution are known to generate vorticity (Truesdell 1952, see Kornriech & Scalo 2000 and references therein). Since the internal structure behind radiative shocks in inhomogeneous media is not clear, we treated the velocity dispersion of the shocked gas as a model parameter, constant for all shocks. In the future, we will explore the effects of letting the the internal shell velocity dispersion depend on the velocity of the ambient gas in the shock's frame of reference as suggested by Elmegreen (1994).

It is assumed that the gas is distributed in a thin sheet perpendicular to the simulated galactic plane and that the radius of curvature of the shell is large compared to its thickness. In the absence of accretion or stretching, the condition for gravitational collapse, and thus star formation, becomes a threshold condition on the shell column density:

$$\frac{\pi G \mu}{c} > \omega_c, \qquad (5)$$

where $\mu$ is the gas column density of the sheet and $c$ is the shell internal velocity dispersion.

The inclusion of accretion onto the slab and expansion of the slab have been studied by Elmegreen (1994) and Comeron & Torra (1994) in the context of a shell expanding into a uniform medium. Appendix B gives a derivation similar to theirs, but which leaves the accretion and expansion rates independent. The action of accretion and expansion do not affect the wavenumber or mass of the fastest growing mode, they just retard the growth rate making collapse more difficult. The condition for significant collapse becomes (see Appendix B)

$$\left(\frac{\pi G \mu}{c}\right)^2 > (\omega_c + \nabla \cdot \mathbf{v_0} + A/\mu)(\omega_c + \nabla \cdot \mathbf{v_0}). \qquad (6)$$

where $\nabla \cdot \mathbf{v_0}$ is the rate of stretching in the shell due to expansion and $A$ is the accretion rate of ambient gas onto the shell. This is the condition for star formation adopted in the simulations. In practice the stretching and accretion terms were found to be relatively unimportant. A physical argument for the slight effect of stretching was given by Whitworth et al. (1994).

Our use of the shell gravitational instability criterion in the simulations does not imply that the shells will fragment at sizes given by Ehlerova et al. (1997) and others for shells expanding into smooth ambient media. This is because in the present simulations star formation occurs at multiple sites within the simulation domain, and the mean separation of these sites is generally small. Thus a given shell in the present simulations expands into an extremely inhomogeneous density and velocity field, and its evolution is dominated by interactions with the other shells of various ages, which have in turn been deformed by previous interactions. In addition, new shells propagate, in part, within their parent shells, whose densities can be large relative to the average density. In contrast, the model shells of Ehlerova et al. (1997) and similar papers expand into a smooth medium and only become unstable when their sizes are of order 1 kpc.

Prescriptions for SF that have been used in previous work on kpc-scale galaxy evolution, protogalactic collapse, galaxy interactions, isolated disk (and LSB disk) galaxies and dwarf galaxies usually involve either some Schmidt-like assumption about the density dependence of the SFR, (e.g. Chiang & Prendergast 1985, Theis et al. 1992, Mihos & Hernquist 1994, Anderson & Burkert 1988), or some combination of constraints including sound crossing time exceeds free-fall time (a Jeans criterion) negative local velocity divergence (convergent flow), or density exceeding a constant threshold density (usually to imitate a condition that the cooling time be much smaller than the dynamical time). Examples are Katz, Weinberg & Hernquist (1996), Vazquez-Semadeni, Passot & Poquet (1996), Steinmetz & Muller (1995), Gerritsen (1997), Lia, Casrraro & Chiosi (1998), Navarro & Steinmetz (1997), Mori et al. (1997).

In contrast to all this work, our simulations assume that star formation only occurs in gravitationally unstable sheets or filaments, for which it is essentially a critical column density *through* the sheet or filament that controls local gravitational instability by eq. 6, for a given prescribed and constant gas velocity dispersion within the sheet or filament (since effects due to shell dilatation and accretion were found to be negligible). The major cpu bottleneck in this prescription is that a filament-finding algorithm must be used in order to check all the column densities (sec. 2.3.1 below). Major simplifications are that the velocity dispersion in filaments is a constant, so the threshold column density is a constant, and that the timescale for instability is given by the growth timescale for the fastest-growing mode, also assumed constant. Because we assume that the flows are advection-dominated with no pressure, there is no energy equation or cooling condition to account for. These reductions in the level of modeling reduce the number of free parameters, allow a more "transparent" attempt to understand the physical phenomena which result (which turn out to be complex enough without including additional processes), yet contain a physically-reasonable representation of a mode of star formation that probably occurs in real galaxies (see the resent summary of references in Elmegreen 1999).

The total mass of stars formed in an unstable region is assumed to be given by $M_{\rm stars} = \epsilon M_{\rm gas}$, where the star formation efficiency $\epsilon$ is assumed constant. The gas mass $M_{gas}$ is given by the product of the gas column density through the shell $\mu_{\rm sh}$ and the vertical cross-sectional area of a simulation cell $h\Delta x$, where $h$ is the assumed scale height of the galaxy. We feel that this estimate of the gas mass is the most physically plausible given our contention that the unstable regions reside in gas shells. An alternative estimate would be to simply use the gas mass in the local star-forming site; however, the resulting masses would be affected by numerical diffusion, which controls the shell's thickness and thus average density. By using the column density, we have essentially eliminated the direct effects of numerical diffusion from the cluster mass estimate.

The Jeans mass is not relevant to this calculation since the desired mass is the total mass of stars formed in the simulation cell and not the mass of an individual condensation. For a 2-D sheet, the Jeans mass is simply $M_J = \mu\lambda_J^2$, where $\mu$ is the shell surface density and $\lambda_J$ is the Jeans length. Since the number of Jeans condensations $N_J$ in a 2D region of size $l$ is $(l/\lambda_J)^2$, the total mass of collapsing gas is $M_J N_J = \mu\lambda_J^2(l/\lambda_J)^2 = \mu l^2$ which is simply the total mass contained in the region and is independent of the Jeans mass. Still, we refer to the newly formed stars in a given cell as a "cluster" with the understanding that it may be several unresolved clusters or a partial cluster. If the simulations are regarded as applying to smaller regions after suitable scaling, for example the interiors of molecular clouds, then "cluster" may be interpreted as "protostar."

By assuming for simplicity a constant power-law IMF with logarithmic slope $\Gamma = -1.5$ and upper and lower mass cutoffs of 0.1 $M_\odot$ and 40 $M_\odot$, we estimate the number of massive stars formed with $M > 8$ $M_\odot$ to be $N_* = M_{\rm stars}/10^3$ $M_\odot$. Substituting the parameter values adopted for the present simulations $\Delta x = 7.8$ pc (see below), $h = 100$ pc, and $\epsilon = 0.1$ (see section 2.5), we find that $N_* = 0.8$ $N_{\rm fil,21}$, where $N_{\rm fil,21}$ is the filament column density in units of $10^{21}$ cm$^{-2}$. This is the number of massive stars which form *per lattice site* along the filament. Thus, if a filament becomes unstable over many lattice sites, the total number of massive stars produced by the filament could be considerably larger.

### 2.3 Filament properties

*2.3.1 Identification*

The identification of filaments, or "shells," in the simulation is required to determine the shell column densities and expansion rates used in the star formation criterion (eq. 6). The technique employed looks for local maxima in the den-sity field along multiple lines of sight. If a given cell is a local maximum along at least two of four possible lines (parallel to the x-axis, y-axis, and diagonals) it is classified as lying on a filament "ridge." Requiring that a cell be a local maximum in all four directions would locate cloud peaks. On the other hand, virtually all points where the density field is convex are local maxima in at least one direction (the direction tangent to the local density contour). This method depends entirely on the coarse-graining of the density field provided by the discreteness of the simulation.

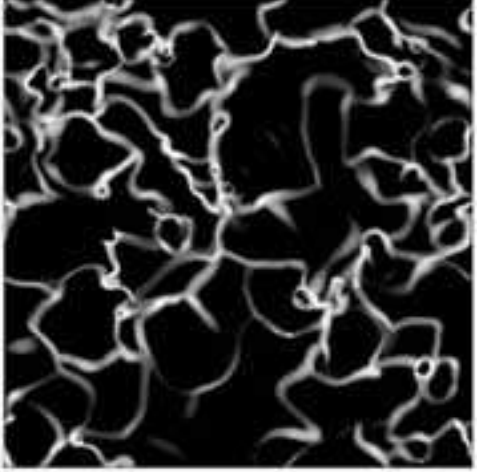

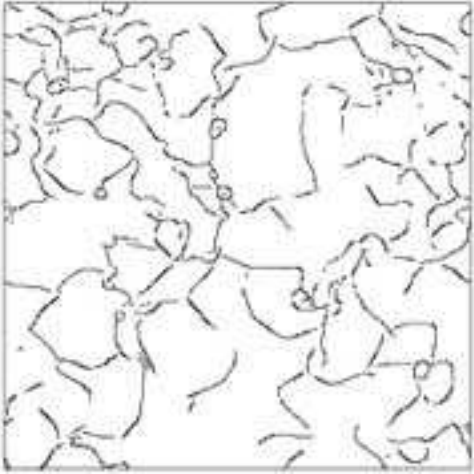

Figure 1- A typical $256^2$ density field from a simulation with stellar forcing (top). Filaments identified by the filament-finding routine (bottom). Plotted are line segments of length equal to the lattice spacing $\Delta x$ at each lattice site determined to lie on a filament ridge. The angle of each segment indicates the estimate of the local filament orientation. The local filament column density is determined by integrating the density distribution along a line perpendicular to the line segments shown.

As an example, we present the application of this technique to an idealized filament oriented along the x-axis with a density gradient along the length of the filament and a linear falloff along the y-axis. The density profile is given by

$$f(x,y) = f_0 + ax - b|y|. \tag{7}$$

If this function is coarse-grained by overlaying boxes of width $\Delta x$, the average density in a box is $\overline{f}(i,j) = \int\int f(x,y) dx dy$, where the integration limits correspond to the sides of the box. If we consider a box centered at $(x,y) = (0,0)$, clearly the boxes above and below it on the y-axis will have average densities less than the center box making it a local maximum along $\hat{y}$. On the other hand, the average density in boxes along the x-axis increase with increasing x, so the local box is not a local maximum along the x-axis. Thus, the classification of this simple example as a filament depends on the average density in the diagonal boxes. It is straightforward to show that the central box will be a maximum along the diagonals, signaling the presence of a filament, if and only if $|a| < 0.75b$. Thus, if the density gradient along the filament $a$ is too steep, the "filament" is not recognized. Similar arguments can be made for idealized filaments with arbitrary orientations and profiles.

Figure 1 shows a section of the density field taken from a typical simulation and the corresponding filament ridges identified by this method. The local orientation of each filament indicated by the directionality of the line segment is determined by the method discussed below. As the figure shows, the identified filament ridges reproduce the structure in the density field well.

### 2.3.2 Shell orientation

The local directionality of a shell is required to estimate its properties such as the column density and expansion rate. Attempts to estimate this direction by fitting linear functions to the local gas density distribution or computing the eigenvectors of the local gas inertia tensor were found to be computationally expensive and often gave erroneous results due to confusion with neighboring filaments. We found that estimating the subgrid position of the filament ridge based on the gas density at the neighboring lattice sites, and least-square-fitting a line to those neighboring positions, gave satisfactory results with low computational overhead.

The method is as follows. For each direction in which the local lattice site is found to be a local maximum, the position of the local maximum along that line is estimated through parabolic interpolation. The maximum along a give line is given by:

$$\vec{x} \approx \vec{x_0} - \hat{n}\frac{\Delta x}{2} \frac{\rho(\vec{x_0} + \hat{n}) - \rho(\vec{x_0} - \hat{n})}{\rho(\vec{x_0} + \hat{n}) + \rho(\vec{x_0} - \hat{n}) - 2\rho(\vec{x_0})} \tag{8}$$

where $\hat{n}$ is a unit vector defining the line of interest. The position of the filament ridge is then defined by the average of these position estimates and the local filament orientation is obtained by least-square-fitting a line to the local and neighboring shell position estimates. We label the directions parallel to and perpendicular to this line by unit vectors $\hat{n}_\parallel$ and $\hat{n}_\perp$ for the discussion in the next two sections. Application of this method to test density rings with a range of widths and radii produced errors in the estimate of the position angle less than $5°$.

### 2.3.3 Column densities

The column density of the shell is estimated by integrating the density distribution along a line perpendicular to the shell (i.e. parallel to $\hat{n}_\perp$). The integration is initiated at the shell maximum and proceeds down both sides of the shell until local minima in the density are encountered or a maximum integration length is reached. The planar approximating functions used in the donor-cell advection scheme are used in the integration. Thus, the shell column density may be expressed as

$$N(\vec{x}) = \frac{1}{m_{\rm p}} \int \rho(\vec{x} + s\hat{n}_\perp)\, {\rm d}s \qquad (9)$$

where $\rho(\vec{x})$ is given by Eq. 1 and $m_{\rm p}$ is the average particle mass. For most of the simulations, the maximum integration distance was set to $l_{\rm max} = 10\Delta x$. Since the integration limits depend on the density structure in the simulation, the implementation of the algorithm which calculates the column densities cannot be fully vectorized and thus is relatively expensive. The requirement of a maximum distance in the column density integration serves to reduce computation time. By running several simulations with varying $l_{\rm max}$, we found that this limit does not affect our results.

### 2.3.4 Shell expansion rate

The time scale for the shell stretching is approximately $\tau_{\rm str} = [\nabla_\parallel \cdot \vec{v}]^{-1}$, where $\nabla_\parallel$ represents the gradient of the component of the velocity field along the shell (parallel to $\hat{n}_\parallel$). This gradient operator is a directional derivative and is defined as

$$\nabla_\parallel \cdot \vec{v} \equiv \frac{\partial}{\partial s}[\hat{n}_\parallel \cdot \vec{v}(\vec{x} + s\hat{n}_\parallel)|_{s=0}]. \qquad (10)$$

We estimate this derivative using a first order difference method

$$\nabla_\parallel \cdot \vec{v} \approx \frac{v_\parallel(s + \Delta s) - v_\parallel(s - \Delta s)}{2\Delta s} \qquad (11)$$

where $v_\parallel \equiv \hat{n}_\parallel \cdot \vec{v}$. Setting $\Delta s = \Delta x$ and using a linear interpolation scheme to estimate the velocities, the following expression for the directional derivative results

$$\begin{aligned}\nabla_\parallel \cdot \vec{v} &\approx n_{\rm x}n_{\rm y}(v_\parallel^{i+1,j+1} - v_\parallel^{i-1,j-1}) + \\ &\quad (1-n_{\rm x})n_{\rm y}(v_\parallel^{i,j+1} - v_\parallel^{i,j-1}) + \\ &\quad n_{\rm x}(1-n_{\rm y})(v_\parallel^{i+1,j} - v_\parallel^{i-1,j})\end{aligned} \qquad (12)$$

for $n_{\rm x}n_{\rm y} \geq 0$ and

$$\begin{aligned}\nabla_\parallel \cdot \vec{v} &\approx n_{\rm x}n_{\rm y}(v_\parallel^{i-1,j+1} - v_\parallel^{i+1,j-1}) + \\ &\quad (1-n_{\rm x})n_{\rm y}(v_\parallel^{i,j+1} - v_\parallel^{i,j-1}) + \\ &\quad n_{\rm x}(1-n_{\rm y})(v_\parallel^{i-1,j} - v_\parallel^{i+1,j})\end{aligned} \qquad (13)$$

for $n_{\rm x}n_{\rm y} < 0$ where we have dropped the "$\parallel$" symbol from the unit vector components for notational convenience.

## 2.4 Stellar energy injection

Supernovae provide the primary energy source by which stars drive bulk motions of the interstellar gas (Abbott 1982, Castor 1993). Due to the correlated positions of supernovae in a cluster, the net effect is a "cluster wind." Leitherer et al. (1993) showed that the net power and momentum of the combined winds and supernovae ejecta of a star cluster remain roughly constant over the first $10^7$ years and declines afterward (see, however, Oey & Smedley 1998). They found this result to be largely independent of the the model parameters including the metallicity and the detailed form of the IMF.

We model the energy injection from a star cluster as a cluster wind (conceptually viewed as the product of multiple supernova events) with constant momentum and power which lasts $\tau_{\rm w}$ years. The total energy released over the lifetime of the cluster is simply $E_{\rm tot} = N_* E_{\rm SN}$, where $N_*$ is the number of massive stars in the cluster which become supernovae and $E_{\rm SN}$ is the energy per supernova. We assume that the transfer of the cluster wind energy into mechanical kinetic energy of the gas within the local star-forming lattice site has an efficiency $f_{\rm mech}$. This efficiency parameter is introduced to parameterize the effects of the unresolved physics related to heating and radiative losses during the formation of the expanding shell. Thus, the kinetic energy of the local gas is given by $E_w = f_{\rm mech} E_{\rm tot}$ and its average velocity is $v_w = (2E_w/m_{\rm cell})^{1/2}$, where $m_{\rm cell}$ is the mass in the local star forming cell. Substituting the above expression for $E_{\rm tot}$, the expression for the gas velocity becomes

$$v_w = \left(\frac{2f_{\rm mech} N_* E_{\rm SN}}{m_{\rm cell}}\right)^{1/2} \qquad (14)$$

By substituting the expression for the number of massive stars formed in a cluster (see section 2.2.1), $N_* = \epsilon m_{\rm cell}/10^3 M_\odot$, we find that typical values of this velocity are given by

$$v_{\rm w} = 40\,{\rm km\,s}^{-1}\left[\left(\frac{\epsilon}{0.1}\right)\left(\frac{f_{\rm mech}}{0.1}\right) E_{\rm SN, 51}\right]^{1/2}. \qquad (15)$$

Thus 40 km sec$^{-1}$ would be the velocity of the gas at a typical star forming forming lattice site if it received a fraction $f_{\rm mech} = 0.1$ of the total energy produced by the supernovae in a typical star cluster. Our standard value of $f_{mech} E_{SN} = 10^{50}$ erg is close to the value found in detailed 1-dimensional simulations of supernova remnants by Thornton et al. (1998). However our wind velocities are much smaller than usually estimated ($\sim 1000$ km s$^{-1}$) because, in our simulations, the momentum input has to accelerate a mass of gas in a cylinder whose height is equal to the adopted scale height. This small wind velocity is, on the other hand, computationally expedient because larger wind velocities would require smaller time steps. Besides, our wind velocity is only applied at distances from the cluster equal to half a grid size (3.9 pc) and usually occurs within high-density regions, so the expected wind speeds should not be as large as 1000 km s$^{-1}$.

Computationally, the cluster wind is modeled by redistributing the mass in the local star forming site into the site's eight nearest neighbors over a time interval $\tau_{\rm w}$. The redistributed mass is given velocity $v_{\rm w}$ directed radially away from the cluster. After the initial mass and momentum deposition, the dynamics are governed by the donor-cell advection prescription.

## 2.5 Shell expansion

As the cluster wind encounters the surrounding gas a compressed layer forms and expands away from the star formation site, conserving momentum. For a uniform ambient gas distribution, shells driven by internal winds are expected to follow the $R_{\rm sh} \propto t^{2/3}$ law predicted by extending Avedisova's momentum-conserving solution (Avedisova 1972) to 2-D. After the cluster winds subside, the shell would evolve as $R_{\rm sh} \propto t^{1/3}$ as in the 2-D zero-pressure snowplow solution.

Neither of these expansion laws are expected to hold for true wind-blown shells. In each case, the two dimensionality of the simulations changes the expansion power-law exponents. The zero-pressure snowplow solution in 3-D is $R_{\rm sh} \propto t^{1/4}$, and Avedisova's 3-D momentum conserving solution has an expansion law of $t^{1/2}$. Also, Weaver et al. (1977) demonstrated that the hot interior of the shells can settle into pressure equilibrium and that this pressure leads to a faster expansion $R_{\rm sh} \propto t^{3/5}$ than Avedisova's solution. The reason that the shells in the present simulations are expected to follow Avedisova's solution and not Weaver et al. is that we do not model pressure effects. However, we do not feel that the precise form of the expansion law will have a qualitative effect on the global statistical properties and organizational behavior in which we are interested. Thus, the added computational expense of solving an energy equation to better reproduce the shell expansion law does not seem warranted. Also, we note that the $t^{2/3}$ scaling expected in the present simulations is intermediate between the 3-D Weaver solution ($t^{0.6}$) and the 2-D analog ($R_{\rm sh} \propto t^{3/4}$) which we derived using simple dimensional analysis arguments.

These scaling solutions only apply to shells expanding into a uniform medium, but in the present simulations the gas becomes very inhomogeneous, and, besides, the shells usually begin their expansion within a "parent shell" and so even the local ambient density field is anisotropic. For these reasons, none of these idealized scaling solutions are realized, except perhaps in a statistical sense at early times. Rather, the shells are often stretched and distorted as they interact with the surrounding gas density and velocity fields.

## 2.6 Normalizing values

Since we are primarily interested in modeling the dynamics of the ISM in disk galaxies, we choose fiducial time and length scales appropriate to that problem. However, we note that through renormalization, the model could also be applied to smaller-scale, lower-momentum interactions driven by low-mass protostellar winds within a molecular cloud, such as the model suggested by Norman & Silk (1980), or to megaparsec-sized scales in which galactic winds provide the momentum forcing, such as the propagating galaxy formation model of Ostriker and Cowie (1981).

We set the lattice spacing to $\Delta x = 7.8\,{\rm pc}$ giving simulation sizes of $L = 1$ kpc, $L = 2$ kpc, and $L = 4$ kpc for the $128^2$, $256^2$, and $512^2$ grids, respectively. This choice is somewhat arbitrary but represents a compromise between choosing a large enough domain to contain many superbubbles (which are typically observed to have sizes from 50 to 100 pc but vary from less that 10 pc to more than a kiloparsec (Heiles 1979, Chu & Kennicutt 1994, Yang et al. 1995)) and one small enough such that the absence of galactic differential rotation in the simulation is not too severe and such that the shells are resolved by many lattice sites. In the solar neighborhood, the shearing time of a kiloparsec-sized region is roughly $\tau_{\rm shear} = L({\rm d}V/{\rm d}R)^{-1} \approx 2 \times 10^8$ years (Mihalas & Binney 1981). In comparison, the kiloparsec crossing time for the fluid motions in the present simulation is $t_0 = 2.5 \times 10^7$ years (see below). While the effects of differential rotation can have important consequences for large-scale fluid dynamics (e.g. Palous et al. 1994 and Elmegreen 1994 study its effects on expanding supershells), we wish to investigate the effects of advection on the shells, decoupled from other processes.

We adopt an average gas density of $n_0 = 1\,{\rm cm}^{-3}$ corresponding to typical values in galaxies on kiloparsec scales and a disk scale height $h = 100$ pc, although one series of models was run with a variety of values of $n_0$. With this choice, the mass in a cell with the average density $n_0$ is $\overline{m}_{\rm cell} = 190\,M_\odot$. The average gas column density $\overline{N}_{\rm cell}$ *in the plane of the simulation* through a cell with density $n_0$ is $\overline{N}_{\rm cell} = n_0 \Delta x = 2.4 \times 10^{19}\,{\rm cm}^{-2} n_0 (\Delta x/7.8{\rm pc})$.

The cluster wind speed provides a natural characteristic velocity for the simulations. In the previous section a computationally appropriate wind velocity was found to be $v_{\rm w} = 40{\rm kms}^{-1}$. We define a characteristic time scale as the kiloparsec crossing time of the wind $t_0 \equiv 1{\rm kpc}/v_0 = 2.5 \times 10^7$ years.

We note that the two-dimensionality of the simulation poses several problems in assigning physical units. First, real wind-blown shells are initially much smaller than the galaxy scale height and expand into a three-dimensional medium. Since the mass in each simulation cell is assumed to correspond to a column of gas with a height equal to the galaxy scale height $h$ and mass $\Delta x^2 h \rho_0$, then the velocities of the expanding shells would be vastly underestimated initially if they are imagined to sweep up all the gas in the local cell (column). However, once the shell becomes comparable in size to the galaxy scale height, the simulation velocities will be in reasonable agreement with true shell velocities since the affected mass is $R_{\rm sh}^2 h \rho_0 \approx h^3 \rho_0$. In addition, as discussed above, the expected scaling of the shell expansion laws depends on the spatial dimension.

## 2.7 Reynolds number

The Reynolds number $Re$ measures the relative importance of the inertial (advective) to viscous or diffusive terms and may be estimated by comparing the time scales associated with each process. The inertial time is simply $\tau_{\rm inertial} = L/v$ where $v$ and $L$ are characteristic velocities and sizes of a region of the flow. The only diffusive terms present arise from numerical diffusion of the donor-cell scheme. Near discontinuities, the numerical diffusion has the form $\nu_2 \partial_x^2$ where $\nu_2 = 0.5(1 - u/u_{\rm g})u/u_{\rm g}$, where $u_{\rm g} = \Delta x/\Delta t$ is the maximum stable grid velocity. The corresponding time scale is $\tau_{\nu,2} \approx L^2/\nu_2 = 2L^2/(1 - u/u_{\rm g})u/u_{\rm g}$. Thus, the Reynolds number is $Re \approx \tau_{\nu,2}/\tau_{\rm inertial} \approx 2L$, where we have justifiably assumed that typical grid velocities are, on average, considerably less than unity (in units of $v_w$). Thus, for $128^2$ and $256^2$ simulations, the Reynolds number is about $Re = 256$ and 512, respectively, indicating that diffusion is relatively unimportant on the largest scales of the simulations. By equating the diffusive and inertial times, we find that diffusion becomes important only on the smallest scales $l_d \approx \Delta x$. The characteristic thicknesses of filaments in our simulations are a few times $\ell_d$.

Away from discontinuities, the numerical diffusion takes the form $-\nu_4 \partial^4/\partial x^4$ with $\nu_4 = [3(1 - (u/u_{\rm g})^3) - 6(1 - u/u_{\rm g})u/u_{\rm g}]u/u_{\rm g}$. In this case, the "super" diffusion time is approximately $\tau_{\rm v,4} \approx l^4/\nu_4 \approx l^4/(3u/u_{\rm g})$ for small velocities. The resulting Reynolds number $Re \approx l^3/3$ is $Re \approx 7 \times 10^5$ and $6 \times 10^6$ for $128^2$ and $256^2$ lattices respectively.

## 2.8 Initial conditions

The velocity field is initialized as a Gaussian random field, meaning that the spatial Fourier modes have random uncorrelated phases with Gaussian-distributed amplitudes dependent only on the norm of the Fourier wavenumber $k = |\vec{k}|$. These conditions guarantee that the probability distribution function of a given mode is initially Gaussian. Two forms of the assumed initial energy spectrum are used: a scale-free spectrum of the form $E(k) \propto k^{-\alpha}$ and an energy spectrum which peaks at a preferred scale $k_0$, $E(k) \propto k^4 \exp(-k^2/k_0^2)$. The standard deviation of the initial velocity field is set near the characteristic velocity $v_0$ to minimize the transient time. The value of the initial variance is found to not affect the post-transient behavior. No stars are initially present. For the majority of the runs, the initial density field is assumed to be uniform. Fluid advection, however, quickly causes density inhomogeneities to develop which reflect the form of the initial velocity fluctuations. This is similar to the manner in which Elmegreen et al. (1995) initialized simulations of shock-turbulence interactions.

## 2.9 Boundary Conditions

Periodic boundary conditions were used for all the simulations presented in this work, so the simulations take place on a torus. Using this choice, the simulation domain is intended to represent a portion of a more extended region of space with statistically similar dynamics. While this choice is widely used in spatial simulations, it can introduce artificial correlations on scales comparable to the domain size.

In the future, we plan on implementing the code on a differentially rotating coordinate system in which concentric rings of lattice sites are allowed to slip past each other to mimic the effects of galactic differential rotation (e.g. Seiden & Gerola 1982, Palous et al. 1994). However at present, our main interest is in isolating the behavior of wind-advection interactions.

## 3 RESULTS: SPATIAL DISTRIBUTION

### 3.1 Model parameters

Table 1 lists the parameter values and a few global statistics for the simulations with star formation. The models are organized such that each series represents the variation of a parameter relative to a standard model (indicated by an asterisk in the table). For runs A, the domain size was varied from $64^2$ to $512^2$ corresponding to linear scales from 0.5 kpc to 4 kpc. We define the standard model as the $256^2$ simulation. Notice that all the statistical quantities listed show little variation for domain sizes larger than $128^2$. As pointed out in section 2.1, the variation of domain size is not equivalent to conventional resolution convergence tests because the resolution is not being varied independently of other physical parameters. However, the stability of the quantities in Table 1 does show that the results are not affected by resolution in the sense of varying the domain size at fixed grid spacing.

In runs C, the parameter varied is the internal gas velocity dispersion of shells $c_{sh}$ which determines the shell star formation column density threshold $N_{sf}$. In the standard model this parameter value is set to $c_{sh} = 1$ km s$^{-1}$ which corresponds to $N_{sf,21} = 1.0$. The models labelled with a dagger refer to runs that use the same parameters as the standard model except that the domain size is $128^2$ rather than $256^2$.

In runs D, the time delay $\tau_d$ between the initiation of local gravitational instability and the onset of the cluster winds was varied, with the value for the standard model being $2.5 \times 10^6$ years. As pointed out by the referee, this adopted standard value is probbly too small. The shell growth is dominated by the lifetimes of the most massive stars (Leitherer et al. 1993, Oey 1996), which is a factor of 1.5 - 2 times larger than the standard time delay value adopted here. If cluster winds can be driven by the collective action of protostellar winds, instead of multiple SNe, the onset of the cluster wind should roughly coincide with the formation of the first stars. However, $\tau_d$ also includes the gravitational collapse time, which depends on the density of the gas from which the cluster forms. From these considerations, we think it is likely that the models listed in Table 1 with $\tau_d = 0.5 - 2 \times 10^7$ yr may be the most realistic. These cases are discussed in sec. 3.2.4 and 4.1.3 below.

In runs E, the average energy released per supernovae $E_{SN}$ was varied around the standard value of $10^{51}$ ergs. Runs F vary the average density $n_0$. Within some of the series, the effects of varying the simulation size and the power-law slope of the initial velocity power spectrum were investigated. We also studied a series of "decay" runs in which star formation was "turned off" to study the organization and kinematics of the gas in the absence of stellar forcing.

Each of the simulations were integrated for 2 Gyrs. Notice that the integration time (about 40 crossing times for the 2 kpc simulations) is much larger than in most previously published kpc-scale simulations, because we were especially concerned with the possibility of long transient times and wanted to search for any relatively large time-scale correlations induced by spatial self-organization. Such long integration times were possible because of the intermediate level of representation of our model (e.g. self-gravity only included as a local instability prescription) and the use of a first-order differencing scheme.

The physical motivations for our choice of normalizing parameters were given in section 2.5. Here, we briefly list these values. We set the star formation efficiency to $\epsilon = 0.1$ and assume that one massive star forms per total 1000 $M_\odot$ of stars formed. An initial average density of $n_0 = 1$ cm$^{-1}$ is chosen for all models except for series F. Assuming that the fraction of the total energy released by supernova which is transferred to kinetic energy of the shell is $f_{mech} = 0.1$ (consistent with Thornton et al. 1998), the characteristic velocity and kiloparsec crossing time of the wind are $v_0 = 40$ km s$^{-1}$ and $t_0 = 2.5 \times 10^7$ years respectively. The duration of a typical simulation is 2 Gyr or 40 grid crossing times for the $256^2$ simulations. Assuming a galactic scale height of 100 pc, the mass in a cell with this average density is $m_{sf} = 1.7 \times 10^4$ $M_\odot$. A flat energy spectrum of the initial velocity field with equal power per logarithmic wavenumber bin was chosen for most of the runs, but some $E_k \propto k^{-2}$ initial power spectra were also studied.

Gas consumption by stars is not included in the models presented. Gas consumption was studied, but was found to simply lead to a gradual decline in the global star formation

**Table 1.** The wind duration is $\tau_w = 10$ Myr for all runs. The columns are (2) linear grid size, (3) time delay in Myrs, (4) average gas density, (5) Energy per supernovae in $10^{51}$ ergs, (6) shell internal velocity dispersion in km s$^{-1}$, (7) logarithmic slope of the energy spectrum of the initial velocity field. Statistics in columns 8 - 18 are based on the last half of the simulations: (8) standard deviation of the density field, (9) average number of density peaks along a line of sight, (10) mass-weighted gas velocity dispersion, (11) velocity dispersion of the shells, (12) average gas kinetic energy, (13) average number of massive stars per kpc$^{-2}$, (14) average number of star clusters per kpc$^{-2}$, (15) average energy injection rate, (16) relative width of the SFR histogram, where $b \equiv N_*/\overline{N_*}$. The ($\dagger$) and ($*$) symbols indicate the standard model at resolutions of $128^2$ and $256^2$ respectively.

| | Parameters | | | | | | Statistics | | | | | | | | |
|---|---|---|---|---|---|---|---|---|---|---|---|---|---|---|---|
| Run (1) | L (2) | $\tau_{d,6}$ (3) | $n_0$ (4) | $E_{SN,51}$ (5) | $c_5$ (6) | $\alpha_0$ (7) | $\sigma(\rho)$ (8) | $N_{los}$ (9) | $\sigma_\rho(v)$ (10) | $\sigma_{f(v)}$ (11) | $\rho v^2$ (12) | $N_*$ (13) | $N_{cl}$ (14) | $L_*$ (15) | $\sigma(b)/b$ (16) |
| Aa | 512 | 2.5 | 1.0 | 1.0 | 1.0 | 0 | 0.211 | 17.5 | 0.124 | 0.0409 | 6.18 | 9.98 | 7.09 | 7.05 | 0.17 |
| b* | 256 | | | | | | 0.210 | 8.7 | 0.131 | 0.0552 | 6.54 | 11.0 | 7.74 | 7.77 | 0.34 |
| c† | 128 | | | | | | 0.208 | 4.3 | 0.122 | 0.0410 | 6.17 | 10.1 | 6.89 | 7.10 | 0.65 |
| d | 64 | | | | | | 0.202 | 2.3 | 0.089 | 0.0353 | 4.15 | 6.4 | 4.36 | 4.52 | 0.94 |
| e | 32 | | | | | | 0.219 | 1.4 | 0.027 | 0.0046 | 0.67 | 1.6 | 0.64 | 1.13 | 3.00 |
| Ca | 128 | 2.5 | 1.0 | 1.0 | 0.3 | 0 | 0.270 | 7.9 | 0.128 | 0.075 | 13.04 | 73.1 | 198.45 | 51.6 | 0.13 |
| b | | | | | 0.5 | | 0.237 | 5.9 | 0.130 | 0.068 | 10.00 | 31.8 | 44.11 | 22.4 | 0.28 |
| c† | | | | | 1.0 | | 0.208 | 4.3 | 0.122 | 0.0410 | 6.17 | 10.1 | 6.89 | 7.10 | 0.65 |
| d | | | | | 1.3 | | 0.195 | 4.1 | 0.130 | 0.0429 | 7.76 | 9.1 | 4.94 | 6.42 | 0.69 |
| e | | | | | 1.5 | | 0.195 | 3.9 | 0.058 | 0.033 | 1.54 | 1.7 | 0.77 | 1.20 | 1.24 |
| f | | | | | 2.0 | | 0.163 | 3.2 | 0.059 | 0.032 | 1.31 | 0.8 | 0.28 | 0.56 | 1.63 |
| g | 128 | 2.5 | 1.0 | 1.0 | 0.3 | -2 | 0.261 | 7.4 | 0.132 | 0.0530 | 12.88 | 60.5 | 139.05 | 42.7 | 0.15 |
| h | | | | | 1.0 | | 0.206 | 4.4 | 0.124 | 0.058 | 6.19 | 10.5 | 7.36 | 7.14 | 0.56 |
| i | | | | | 1.3 | | 0.197 | 4.0 | 0.114 | 0.0433 | 5.40 | 6.6 | 3.58 | 4.66 | 0.76 |
| j | | | | | 1.5 | | 0.188 | 3.7 | 0.120 | 0.0349 | 5.21 | 5.8 | 2.68 | 4.09 | 1.00 |
| k | | | | | 2.0 | | 0.191 | 3.6 | 0.093 | 0.0424 | 2.23 | 2.0 | 0.70 | 1.41 | 1.05 |
| l | 256 | 2.5 | 1.0 | 1.0 | 0.5 | 0 | 0.241 | 12.1 | 0.137 | 0.071 | 10.44 | 32.0 | 44.85 | 22.61 | 0.16 |
| m | | | | | 0.7 | | 0.225 | 10.3 | 0.134 | 0.0657 | 8.60 | 19.2 | 19.33 | 13.57 | 0.22 |
| n* | | | | | 1.0 | | 0.210 | 8.7 | 0.131 | 0.0552 | 6.54 | 11.0 | 7.74 | 7.77 | 0.34 |
| o | | | | | 1.5 | | 0.189 | 6.8 | 0.121 | 0.0518 | 4.48 | 4.8 | 2.25 | 3.39 | 0.53 |
| p | | | | | 2.0 | | 0.179 | 6.4 | 0.127 | 0.0500 | 4.49 | 3.9 | 1.38 | 2.76 | 0.70 |
| q | | | | | 3.0 | | 0.162 | 5.4 | 0.094 | 0.0385 | 2.01 | 1.1 | 0.52 | 0.78 | 1.14 |
| Da | 128 | 0.0 | 1.0 | 1.0 | 1.0 | 0 | 0.213 | 4.5 | 0.087 | 0.0304 | 3.13 | 4.2 | 3.68 | 2.96 | 0.81 |
| b | | 0.5 | | | | | 0.215 | 4.6 | 0.105 | 0.050 | 5.22 | 7.0 | 5.79 | 4.94 | 0.70 |
| c | | 1.0 | | | | | 0.213 | 4.5 | 0.099 | 0.047 | 4.06 | 5.1 | 4.06 | 3.60 | 0.72 |
| d† | | 2.5 | | | | | 0.208 | 4.3 | 0.122 | 0.0410 | 6.17 | 10.1 | 6.89 | 7.10 | 0.65 |
| e | | 5.0 | | | | | 0.196 | 4.1 | 0.167 | 0.079 | 11.38 | 22.8 | 13.18 | 16.1 | 0.41 |
| f | | 10.0 | | | | | 0.191 | 4.0 | 0.202 | 0.090 | 14.48 | 30.7 | 13.42 | 21.7 | 0.41 |
| g | 256 | 0.5 | 10 | 1.0 | 1.0 | 0 | 0.215 | 8.8 | 0.113 | 0.0504 | 5.05 | 7.7 | 6.44 | 5.44 | 0.38 |
| h* | | 2.5 | | | | | 0.210 | 8.7 | 0.131 | 0.0552 | 6.54 | 11.0 | 7.74 | 7.77 | 0.34 |
| i | | 5.0 | | | | | 0.203 | 8.3 | 0.157 | 0.0727 | 9.95 | 17.4 | 10.24 | 12.30 | 0.29 |
| j | | 10.0 | | | | | 0.193 | 7.9 | 0.200 | 0.0855 | 13.93 | 35.0 | 15.20 | 24.73 | 0.29 |
| k | | 20.0 | | | | | 0.176 | 7.5 | 0.228 | 0.0898 | 17.23 | 58.1 | 17.00 | 41.06 | 0.20 |
| Ea | 128 | 2.5 | 1.0 | 0.25 | 1.0 | 0 | 0.206 | 4.7 | 0.062 | 0.028 | 1.51 | 5.1 | 3.50 | 0.90 | 0.78 |
| b | | | | 0.50 | | | 0.206 | 4.5 | 0.086 | 0.040 | 2.90 | 6.5 | 4.46 | 2.30 | 0.66 |
| c† | | | | 1.00 | | | 0.208 | 4.3 | 0.122 | 0.0410 | 6.17 | 10.1 | 6.89 | 7.10 | 0.65 |
| d | | | | 2.00 | | | 0.203 | 4.3 | 0.173 | 0.085 | 13.65 | 15.0 | 10.67 | 21.2 | 0.53 |
| e | | | | 4.00 | | | 0.199 | 4.3 | 0.266 | 0.122 | 29.53 | 23.5 | 16.66 | 66.3 | 0.48 |
| f | 128 | 2.5 | 1.0 | 0.25 | 1.0 | -2 | 0.217 | 4.4 | 0.056 | 0.030 | 1.59 | 4.5 | 2.98 | 0.79 | 0.80 |
| g | | | | 0.50 | | | 0.208 | 4.4 | 0.092 | 0.041 | 3.03 | 6.9 | 4.76 | 2.43 | 0.59 |
| h | | | | 1.00 | | | 0.206 | 4.4 | 0.124 | 0.058 | 6.19 | 10.5 | 7.36 | 7.4 | 0.56 |
| i | | | | 2.00 | | | 0.202 | 4.4 | 0.177 | 0.088 | 14.48 | 15.5 | 10.96 | 21.9 | 0.59 |
| j | | | | 4.00 | | | 0.199 | 4.2 | 0.274 | 0.122 | 34.14 | 28.3 | 19.91 | 79.9 | 0.47 |
| k | 256 | 2.5 | 1.0 | 0.25 | 1.0 | 0 | 0.214 | 8.7 | 0.062 | 0.0276 | 1.35 | 4.35 | 2.99 | 0.77 | 0.42 |
| l | | | | 0.50 | | | 0.214 | 8.7 | 0.089 | 0.0398 | 2.95 | 6.98 | 4.92 | 2.46 | 0.43 |
| m* | | | | 1.00 | | | 0.210 | 8.7 | 0.131 | 0.0552 | 6.54 | 11.0 | 7.74 | 7.77 | 0.34 |
| n | | | | 2.00 | | | 0.207 | 8.4 | 0.192 | 0.0860 | 15.13 | 17.6 | 12.45 | 24.88 | 0.34 |
| o | | | | 4.00 | | | 0.202 | 8.5 | 0.274 | 0.1255 | 29.82 | 24.5 | 17.35 | 69.26 | 0.26 |
| Fa | 256 | 0.0 | 0.33 | 1.0 | 1.0 | 0 | 0.162 | 3.7 | 0.0271 | 0.094 | 0.056 | 0.669 | 0.375 | 0.26 | |
| b | | | 0.67 | | | | 0.194 | 6.3 | 0.0323 | 0.113 | 0.069 | 2.66 | 2.98 | 2.11 | |
| c | | | 1.00 | | | | 0.210 | 8.8 | 0.0411 | 0.127 | 0.090 | 6.76 | 11.5 | 8.12 | |
| d | | | 1.40 | | | | 0.225 | 11.1 | 0.0464 | 0.129 | 0.108 | 11.57 | 26.6 | 18.70 | |
| e | | | 2.00 | | | | 0.240 | 14.1 | – | 0.131 | 0.130 | 20.51 | 62.2 | 43.60 | |

rate and gas velocity dispersion, as expected. Since this decline occurs only over large secular time scales, we turn off stellar gas consumption for the runs presented here in order to produce simulations whose temporal behavior is nearly statistically stationary, so that time averages can be reasonably defined, and so that simulation data samples from different times can be combined to decrease noise in statistical estimates for quantities like the velocity and mass probability distributions. However it is important to keep in mind that results that are found to depend on the average SFR for different models apply to a single model at different times if gas consumption were included, since in that case the average SFR decreases with time. Models including gas consumption are described in Chappell (1997).

Since the filaments in the 2-D simulations are meant to represent expanding and interacting shells in 3-D, we will use the terms "filaments" and "shells" interchangeably in referring to these density structures. However, it should be remembered that in 3-D the morphology will likely be a mixture of shell-like (i.e. two-dimensional) and filament-like (i.e. one-dimensional) structures.

### 3.2 Spatial distributions of the stars and gas

#### 3.2.1 Standard model

Figure 2 shows the logarithm of the gas surface density (grey scale image in each panel) and recent star formation activity (right panels) at two times separated by $7 \times 10^7$ yrs for the standard model. The filled circles, open circles, and crosses indicate the birth positions of clusters with ages between zero and $10^7$ years, $10^7$ and $5 \times 10^7$ years, and $5 \times 10^7$ and $10^8$ years respectively. The clusters in the youngest age bin have active winds while the winds from the oldest clusters shown have long since subsided. The symbols mark the positions of the clusters at the time when their winds were active in order to show how the spatial distribution of the star formation *sites* evolve in time. The change in the cluster positions (the clusters inherit the velocity of their parent gas clouds) between the time of their birth and the time that the density field is shown is typically less than a lattice spacing for the youngest clusters, but can range up to a few hundred parsecs for the oldest ($10^8$ year) clusters.

The density fields shown in Fig. 2 exhibit a network or web of shells or filaments which are distributed nonuniformly over the grid. Large "voids" and concentrations of shells can be seen at both times. The gas distribution qualitatively resembles the aperture synthesis H I map of the LMC presented by Kim et al. (1998). The most recent star formation events (filled circles) generally trace the large-scale gas distribution. However, on small scales, winds from all but the very youngest clusters have excavated low-density cavities around the clusters and initiated expanding gas shells (see, for example, the clusters and surrounding shells in the boxed region at time $t = 1.99$ Gyrs.) We find that the fate of these shells depends largely on the structure of the surrounding gas density and velocity fields. In some instances the shell merges with the ambient gas but never reaches the star formation threshold, and thus does not directly trigger the formation of more clusters. In other cases, shell mergers lead to the formation of lone clusters as evidenced by the presence of isolated clusters in the right hand panels. Much more common, however, is the tendency for star clusters to form in large groupings.

Figure 3 shows the time evolution of the boxed region of the density field between the two times shown in Fig. 2. In the lower region of the the initial panel, two young star clusters have just initiated two expanding shells. These shells collide at $\Delta t = 10$Myrs, triggering the formation of two new clusters. By $\Delta t = 30$ Myrs, the shells have merged to form a single "supershell" which is expanding toward the filament in the center of the panel. At time $\Delta t = 40$ Myrs, the merging of these two density structures triggers a new round of star formation in the center of the panel and in the lower left corner. The expanding shells from these two star formation events leads to a cascade of multiple propagating star formation events, until, at time $\Delta t = 70$ Myrs, a series of nested shells is apparent. The organization of propagation events in this model is controlled to a large degree by the surrounding gas distribution, which is determined by older propagation events. Thus, propagating star formation in this model does not evolve as a uniform progression through a cloud as originally suggested in the "burning cigar" model of Elmegreen & Lada (1977), nor does it lend itself to descriptions involving the instabilities of expanding shells in a uniform medium (Comeron & Torra 1994, Elmegreen 1994). Rather, even on the relatively small scales of individual propagating events, the process appears as a complex cascade mediated by multiple interacting shells. The specific ionization-shock fron model of Elmegreen & Lada (1977) is of course not refuted by the present calculations, since their model involves scales which we cannot resolve; we only contrast our result with their model in a generic sense.

The emergence of large-scale organization in the star formation activity is apparent in Fig. 2. The large coherent groups of stellar clusters is unexpected. These groups stimulate new star formation in neighboring regions, causing the locus of star formation activity to "wander" throughout the simulation. However, this apparent motion is complex in the sense that it depends on the organization of the ambient gas, which evolves through the action of fluid advection and the collective action of many stellar generations. Thus, the organization and progression of the large-scale star formation activity appear not to occur as a single propagating star formation wave in which stars form consecutively behind an advancing front, but rather as the organization of many smaller-scale propagation events which interact in non-trivial ways.

For example, the winds from the old group of clusters indicated by the x's in the lower-right corner of Fig. 2 at time $t = 1.92$ Gyrs has created a large, low density cavity. The more recent star formation activity indicated by the closed and open circles to the lower left of the boxed region lies at the left edge of this cavity. The "triggering" of this more recent star formation region was a result of the shells driven by the older star clusters at the the center of the this cavity as well as the clusters to the upper left and at the top of the grid.

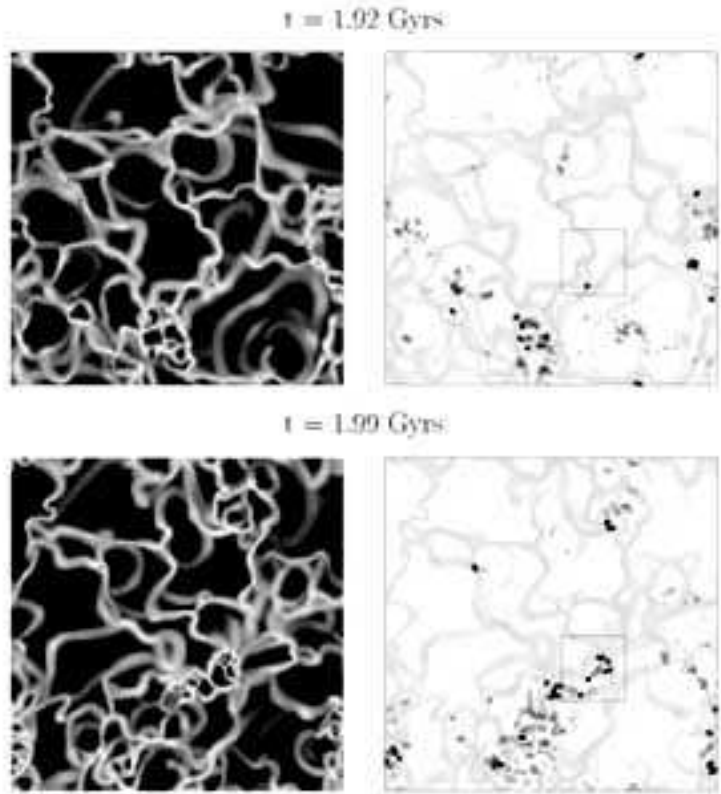

Figure 2. Logarithm of the gas density (left panels) and recent star formation activity (right panels) at two times separated by $7 \times 10^7$ yr for the standard model. Filled circles, open circles, crosses: birth positions of clusters with ages between zero and $10^7$ yr, $10^7$ and $5 \times 10^7$ yr, and $5 \times 10^7$ and $10^8$ yr, respectively.

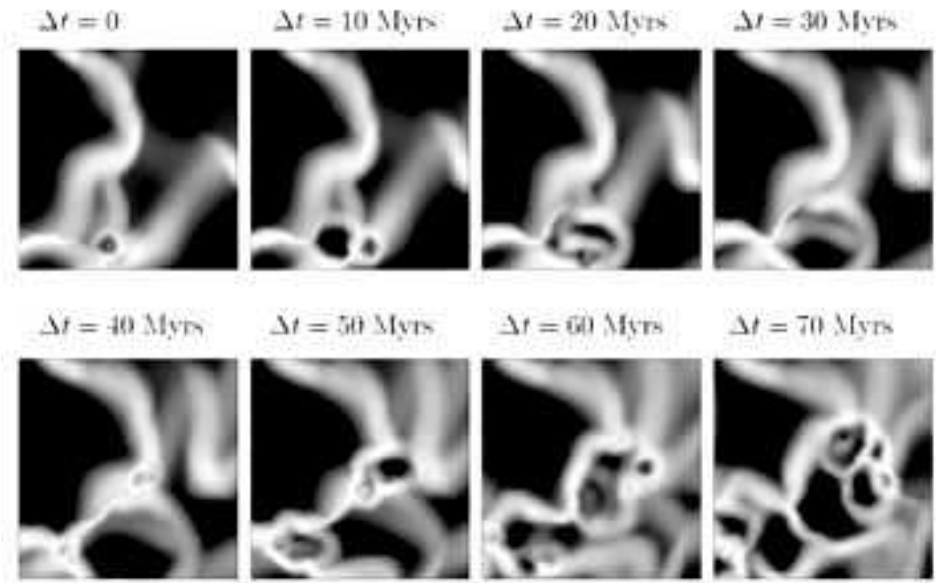

Figure 3. Time evolution of the density field for the boxed region indicated in Fig. 2. The sequence begins at time t=1.92 Gyr (top panel of Fig. 2) and progresses in increments of 10 Myr. The final panel at t=1.99 Gyr corresponds to the lower panels of Fig. 2.

The two-point correlation functions of stars in the simulations are discussed elsewhere (Scalo & Chappell 1998a). Their form is generally power-law, but the logarithmic slope depends on the age of the stars selected and on the average level of star formation in the models.

### 3.2.2 Decay runs

The filamentary structure of the gas in these simulations is a result of advection in the presence of high compressibility. Even in the absence of forcing by cluster winds, initial fluctuations in the gas velocity field lead to filamentary structure. Figure 4 shows the evolution of the density fields for two $256^2$ simulations in the absence of star formation. The initial energy spectra of the left and right panels are of the form $E_k \propto k^4 \exp(-k^2/k_0^2)$ peaking at $k_0 = 8\ L^{-1}$ and $E_k \propto k$, respectively. At early times, soon after shock formation, the density field reflects the structure of the initial velocity field, exhibiting filamentary structures of advected gas at the shock fronts with the spatial wavenumber of the filaments being approximately equal to the initial wavenumber $k_0$ in the left panels. As the shocks collide, they conserve mass and momentum and merge due to the high compressibility of the gas. Thus, the number of shocks (and equivalently filaments) decreases with time and the correlation length of the density and velocity fields increase. The intersection of oblique filaments creates density clumps. These clumps are similar in nature to the clumps that form at the intersections of filaments in large-scale cosmological simulations. At late times, only a few dense clumps and filaments remain. While the filamentary structure of the gas is a direct consequence of the high compressibility of the simulations, comparison of Fig. 2 and Fig. 4 clearly shows that star formation feedback influences the organization of the filaments.

We find that the kinetic energy decays as $t^{-1}$ for all the decay runs that have an initial energy spectrum peaked at a length scale $k_o^{-1} \ll L$, so that many shock interactions occur. This decay looks similar to the results found in more detailed 3D fully hydrodynamic and MHD simulations (MacLow et al. 1998, Stone et al. 1998). This similarity suggests that the insensitivity of the 3D decay to the strength of the magnetic field or the Mach number can be understood as the result of the dominance of fluid advection in controlling the dynamics. A detailed discussion of the kinetic energy decay and power spectrum of the decay simulations and comparison with various analytic results for Burgers turbulence is presented elsewhere, since our major theme here is the influence of star formation. The probability distribution (pdf) of the density field for the decay simulations was presented in Scalo et al. (1998), where it was shown that this density pdf is a power law at large densities because of the effective polytropic exponent being less than unity (zero here), in contrast to isothermal simulations, which are in a sense a singular case, giving lognormal density pdfs. The reader is referred to Scalo et al. (1998) and Passot & Vazquez-Semadeni (1999) for details.

### 3.2.3 Star-formation threshold

The gas column density threshold for star formation in this series of models (series C in Table 1) is found to control both the global star formation rate and the spatial organization of the stars and filaments. Figure 5 shows the organization of the gas filaments (left) and star formation sites (right) for three values of the star formation threshold. As the threshold inceases, the number of filaments and the spatial density of star clusters decreases and the characteristic size of the voids between filament increases. In the upper panels which show the largest values of the star formation threshold, only a few young clusters with active winds are prsent at any given time. The shells expand to large radii before interacting and the characteristic time for interactions is correspondingly long. Much of the shell's life is spent in a state of free expansion in which the driving winds have long since subsided. At the lowest star formation threshold, young shells rapidly interact, triggering vigorous star formation activity and new expanding shells. The characteristic scale of recognizable shells is much smaller and they often interact while still being driven by stellar winds.

The shell size is largely controlled by mass conservation. The threshold parameter sets the upper limit on the gas column density of shells; above that vaue, star formation is triggered and the shell is disrupted by winds from the young embedded stars. For a shell expanding into a uniform medium, the higher threshold column density is reached at a larger shell radius and at a later time than for a lower threshold value. In the present simulations in which most of the mass resides in shells, mass conservation requires that if each filament is on the average more massive, then their number density must be lower. Thus, as the star formation threshold is increased, the characteristic time between successive propagation events decreases and the shells are on average larger and older.

At the largest star formation rates investigated in this series, the gas is highly agitated and the filaments appear shredded. Thus, a typical fluid parcel is not able to decelerate significantly before being swept-up in another shell. Fluid parcels are frequently driven by more than one wind source in these simulations.

### 3.2.4 Time delay

As discussed in section 2, once the gas in a simulation cell satisfies the star formation criterion, a fraction (90% in the present simulations) of the gas in that cell is removed from the grid to form a bound cloud which no longer interacts with the surrounding gas flow. This cloud inherits the space velocity of the gas in the local cell and moves as a collisionless particle. The time delay parameter $\tau_d$ studied in this section is the time between the onset of the collapse of the gas (and formation of the bound cloud) and the onset of stellar winds. Over the lifetime of the cluster, the mass of the bound cloud is returned to the ISM as the cluster wind. Thus, larger time delays $\tau_d$ allow the gas around newly formed proto-clusters to evolve longer before the onset of the cluster winds.

The density distribution and star formation sites for a model with time delay $\tau_d = 10$ Myr is shown in Fig. 6 for three times. As the time delay has been increased relative to the standard model, the cluster formation rate is larger. Accompanying the increased SFR is a higher presence of "chains" of star clusters (compare Fig. 6 where $\tau_d = 10$ Myrs with the standard model in Figure 2 where $\tau_d = 2.5$ Myrs). These stellar chains trace the underlying gas distribution of star-forming filaments.

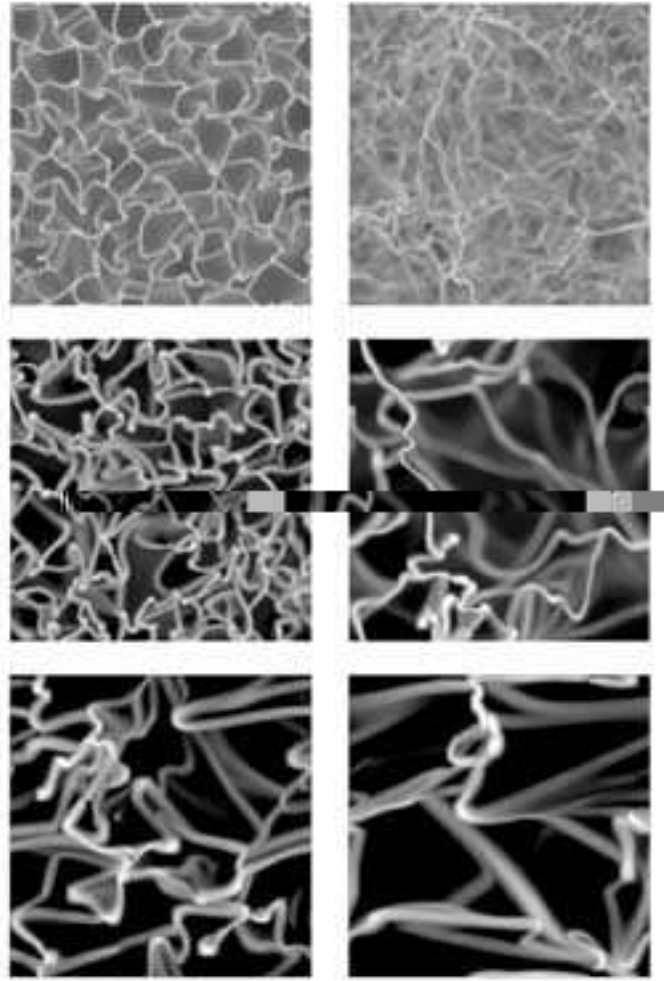

Figure 4. Evolution of the density field for two $256^2$ simulations in the absence of star formation ("decay runs"). *Left panels*: Initial energy spectrum $E_k \sim k\ exp(-k^2/k_o^2)$ with $k_o = 8L^{-1}$. *Right panels*: Scale free initial energy spectrum $E_k \sim k$.

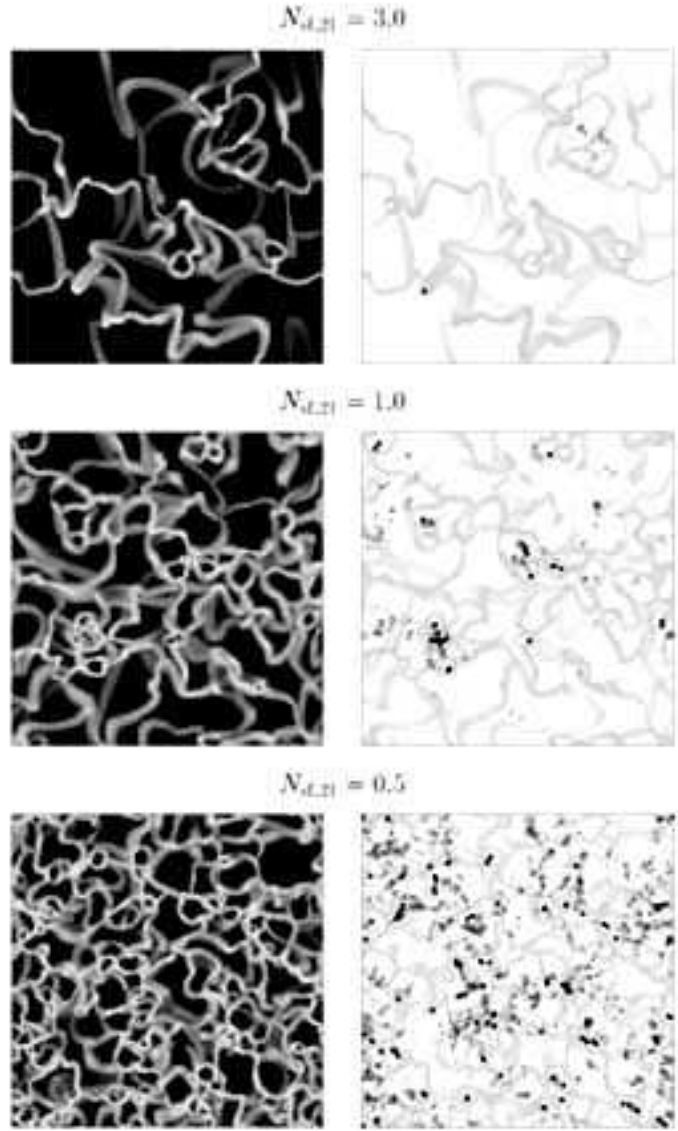

Figure 5. Typical snapshots of gas density (left) and birthsite (right) distributions for simulations in which the column density threshold for star formation is decreased (from top to bottom). $N_{sf,21}$ is the critical filament column density at which star formation was assumed to occur, in units of $10^{21} cm^{-2}$.

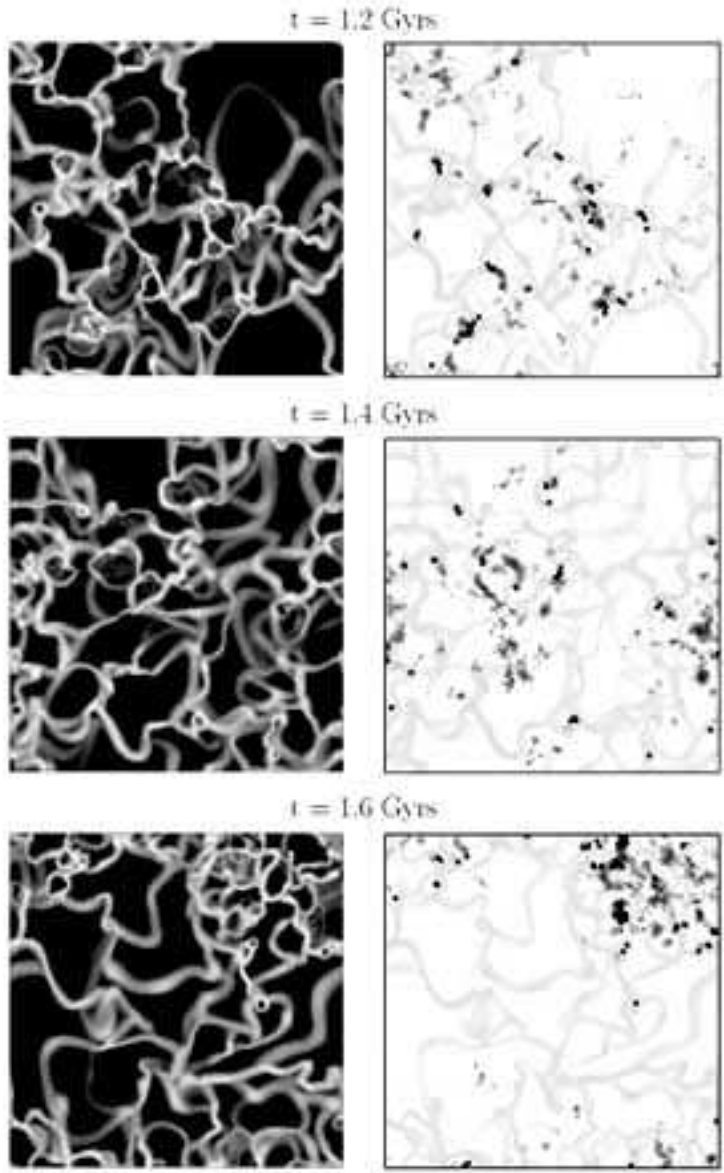

Figure 6. Gas density distribution (left) and star formation sites (right) at three times for a model in which the time delay is 10 Myr, compared to 2.5 Myr for the standard model (cf. Fig. 2 and middle panels of Fig. 5). Overall star formation rate and clustering level are enhanced.

Since the winds from the young stars in a filament disrupt the gas distribution in the filament, a larger time delay increases the likelihood that the column density of a filament, which is sweeping up ambient gas, will reach the star formation threshold before it is disrupted. Thus, more star clusters will form on average in a given filament. For similar reasons, the longer time delays increase the chance that, when the winds from a young cluster compress the gas along the filament, the star formation column density threshold will be reached sooner leading to more star clusters in a given filament. In other words, longer time delays allow filaments to accrete more mass (before being disrupted by the winds from the first stellar cluster to form), which increases the probability that more stars clusters will form "spontaneously" along the filament and which decreases the length scale along the filament for "triggered" star formation as the stellar winds compress the gas along the filament. This scenario also explains the higher global SFR since each star-forming filament produces more star clusters on average.

Clearly, the regularity of these star cluster chains derives from the uniform gas distribution along the filaments. The inclusion of subgrid processes would undoubtedly generate finer-scale structure within the filaments which would lead to less regular stellar distributions. However, our qualitative finding that longer delays between star formation and stellar energy injection produce a higher global SFR with larger spatio-temporal correlations should remain valid.

If the cluster winds are driven by the collective action of the stellar winds, then small time delays may be appropirate, since stellar winds set in during the protostellar phase. However, if the cluster winds are driven by collective SNe, as in the standard picture, the longer time delays are appropriate, since the evoluultionary time required to produce a SN varies from about $3 \times 10^6$ yr for the most massive stars to about $3 \times 10^7$ yr for the least massive stars believed to be capable of SN explosions. Overall, we believe that the long time delay simulations are most likely to represent the real situation, although only a few such simulations are presented here.

*3.2.5 Discussion*

The filamentary structure, observed in this simulation as a result of high compressibility and forcing by stellar winds, is found in many other simulations of the ISM in which stars drive gas motions. The simulations of Chiang & Prendergast (1985) and Chiang & Bregman (1988), which treat the stars as a continuous fluid that are born at a rate proportional to some power of the local gas density and include parameterized power-law cooling and stellar heating, develop a network of filaments near pressure equilibrium. Due to the regularity of these structures, they predict that in three dimensions the voids would be roughly spherical in shape and that the filaments seen in their 2-D simulation would correspond to shells or sheets. More recent work by Rosen et al. (1993) extended their approach by using an integration scheme with less numerical diffusion and found that the resulting filamentary network is less symmetrical and that filaments with a range of densities and temperatures emerge.

The star formation-powered turbulence simulations of Vazquez-Semadeni et al. (1994), Passot et al. (1996) and Vazquez-Semadeni et al. (1997), which include more physics (self-gravity, magnetic fields, expicit heating and cooling, rotation) also exhibit filamentary structures, although they are transient and less regular than the structures found by Chiang & Prendergast (1985). These structures tend to form at the interface between colliding gas streams and lead to star formation, similar to the picture suggested by Elmegreen (1993). Similar structure is found in the isothermal, non-self-gravitating, randomly forced MHD simulations of Padoan et al. (1998). Given the ubiquity of these structures in simulations that involve very different formulations of the physics (i.e. cooling, pressure, star formation "laws", self-gravity, etc.), the present simulations, which only incorporate highly compressible advection and driving by star formation, suggest that these latter physical effects are responsible for most of the filamentary structure seen in these simulations.

Finally, we consider what the gas distribution would look like were it observed with non-perfect resolution. The gray-scale image in Fig. 7 shows the density field convolved with a Gaussian with FWHM of 250 pc. This linear resolution corresponds to angular resolutions of $30'$, $26''$, and $5''$ at distances of 50 kpc, 2 Mpc, and 10 Mpc respectively. Notice also that because the structure is severely underresolved at the resolution of the simulated observations, the gas density actually appears to be distributed in roundish "clouds"; comparison to panel a suggests how misleading conceptual models based on the under-resolved observations might be.

## 4 EVOLUTION OF GLOBAL STATISTICS

### 4.1 Global star formation rate

*4.1.1 Standard model and effects of domain size*

Figure 8 shows the evolution of the global star formation rate (left-hand axis) and gas velocity dispersion (right-hand axis) for the standard model with the domain size ranging from 250 pc to 2 kpc. The star formation rate is defined as the number of massive stars with active winds per kpc$^2$. The time-averaged value of the number of clusters with active winds per unit area is around 7 kpc$^{-2}$ and is largely independent of the simulation size except for the smallest grid. Since the wind duration is $\tau_w = 10$ Myr, this cluster surface density corresponds to a cluster formation rate of $7 \times 10^{-7}$ kpc$^{-2}$ yr$^{-1}$ which is about a factor of three greater than the cluster formation rate for the local Milky Way reported by Elmegreen & Clemens (1985) of $2.5 \times 10^{-7}$ kpc$^{-2}$ yr$^{-1}$. We consider this to be in acceptable agreement with observations, considering that the aim of this work is not to reproduce the detailed properties of any single galaxy, but to investigate the dynamical and organizational aspects of galaxies.

The velocity dispersion shown in the same figure is defined as the rms velocity component weighted by the gas density $\sigma_\rho^2(v_x) \equiv \frac{1}{N} \sum_i (v_i w_i - \overline{v_i w_i})^2$, where $w_i = \rho_i/(\frac{1}{N}\Sigma \rho_i)$ and $N$ is the number of lattice sites in the grid. Thus, it is meant to approximate the width of an optically thin spectral line if the simulation were viewed by an observer in the grid plane with a beam which contains the entire lattice. The average gas velocity dispersion weighted by the density field is around $\sigma_\rho = 0.12 v_0$ which translates to 5 km s$^{-1}$. The velocity dispersion of the filaments for the runs in series A correspond to 2.2 km s$^{-1}$. Stark & Brand (1989) find that the cloud velocity dispersion in our galaxy is about 7 km s$^{-1}$ for cloud masses up to $3 \times 10^5$ M$_\odot$ and declines to 3 km s$^{-1}$ for larger clouds. Again, we consider these results to be in rough agreement.

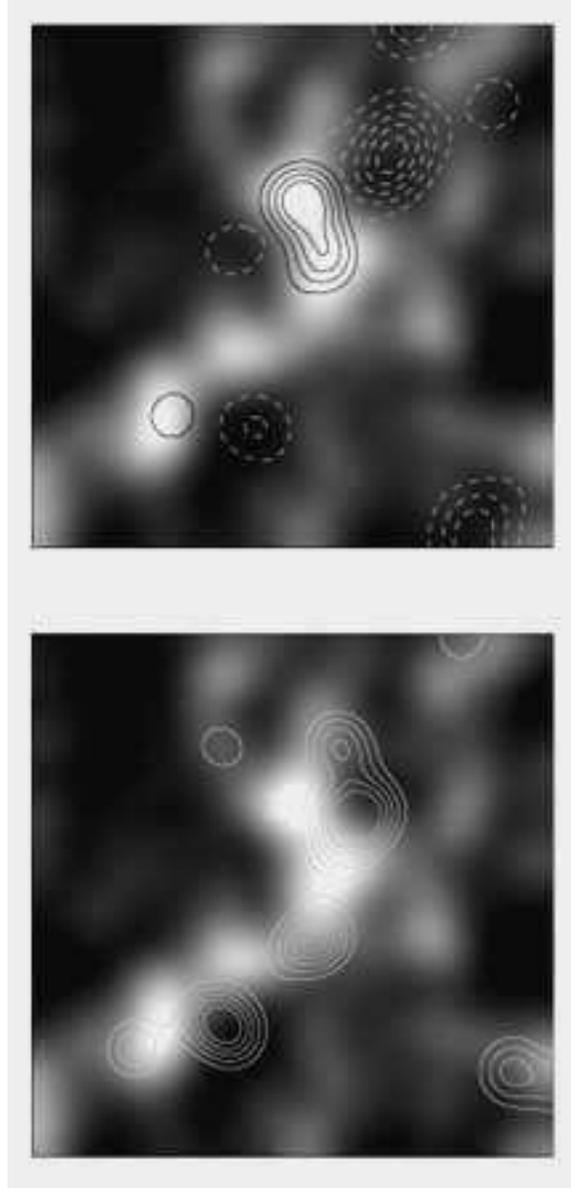

Figure 7. Smoothed gas density and stellar fields for a typical time in a series C simulation. Grey scale images represent gas density smeared with a Gaussian filter and are identical in the two panels. White corresponds to highest densities. Top panel: solid contours are smoothed distribution of clusters with active winds ($t < 10^7$ yr) and dashed white lines show distribution for clusters with ages $5 \times 10^7 < t < 10^8$ yr. Contours in lower panel are distribution of intermediate-aged stars with $10^7 < t < 5 \times 10^7$ yr.

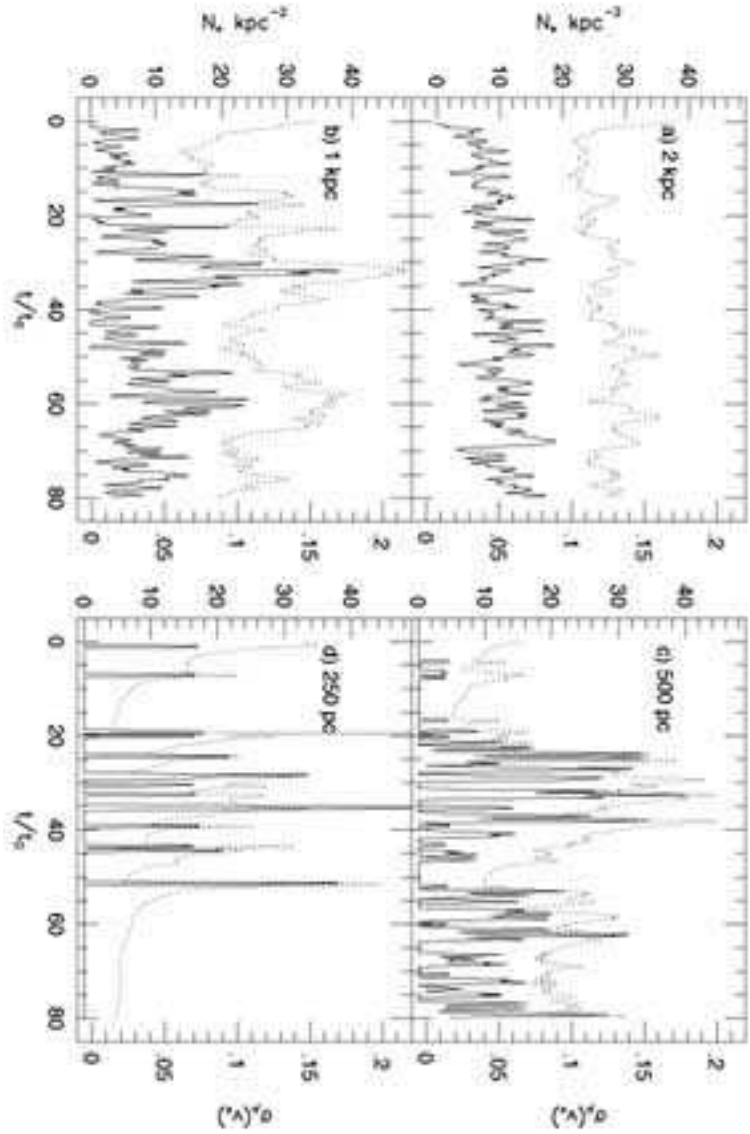

Figure 8. Global star formation rate (number of massive stars per square kpc) and gas velocity dispersion (solid and dashed lines, respectively) as a function of time for models in series A. The simulation sizes are (a) $512^2$, (b) $256^2$, (c) $128^2$, (d) $64^2$.

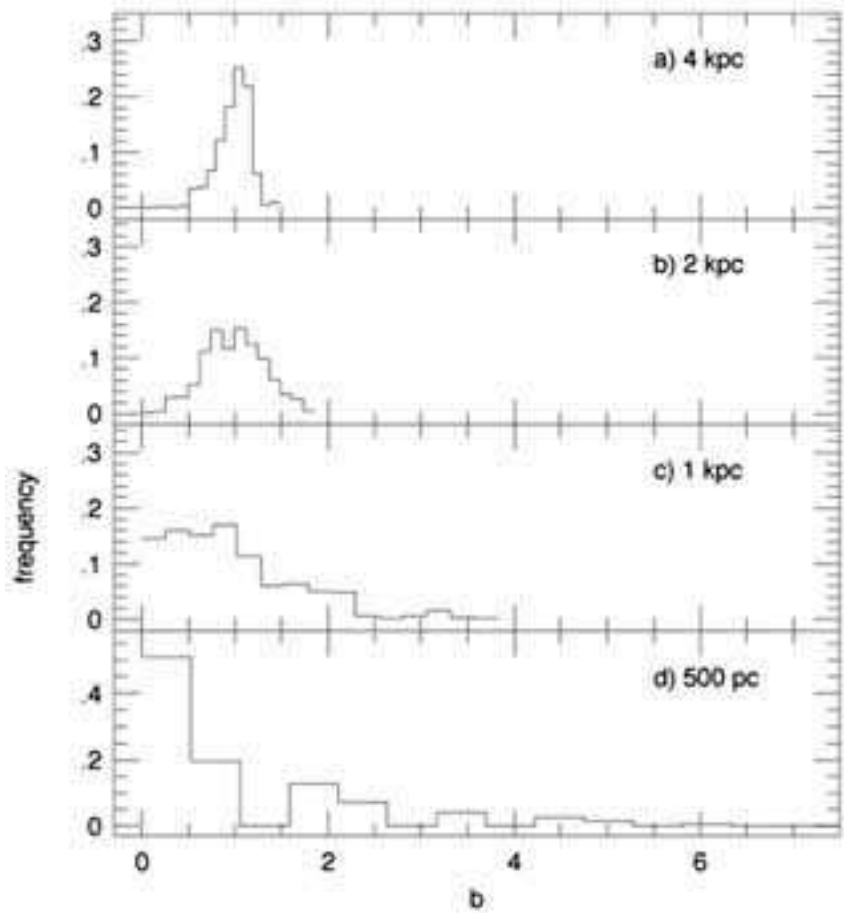

Figure 9. Frequency distribution of the ratio of 0h? present SFR to the mean SFR. For the last Gyr of the simulation, denoted by b, for the four simulations with different spatial extent corresponding to Fig. 8. The burstier behavior with decreasing size is manifested as a broadening of the b-distribution.

Fluctuations in the global statistics increase in amplitude as the system size is decreased. For $L$ less than a kiloparsec, the star formation rate becomes increasingly bursty with long temporal correlations present, while for $L$ greater than a kiloparsec, the star formation rate and gas velocity dispersion settle into a near equilibrium state with no secular variations in time. This transition from continuous star formation to bursty behavior is reminiscent of the transition proposed by Gerola et al. (1980) in their stochastic self-propagating star formation (SSPSF, see Seiden & Gerola 1982 for a review) theory of bursting dwarf galaxies. In both cases, large simulations, in which many regions of propagating star formation are present, exhibit near equilibrium global SFRs with only small fluctuations. However, the physical mechanisms are different. In the SSPSF simulations, bursts occur because percolation of star formation can more easily fill the simulation grid for small sizes. In the present simulations, the lulls of star formation in the small simulations occur because only a few density structures exist at a given time that are near the star formation threshold. Since star formation disrupts the local gas structure, the shell fragments must re-coalesce before another star formation event can occur. Thus a "refractory time" is important in both models, but percolation plays no role in the present models.

Figure 9 shows histograms of the star formation rate normalized to its mean for the last Gyr of the simulation. Observationally, it is equivalent to the histogram of the stellar birthrate parameter $b$, defined as the ratio of the current to past average SFR. Since the simulations studied in the present work are roughly statistically stationary, the histogram of the SFR approximates the histogram of the birthrate parameter b. Histograms of this birthrate parameter have been constructed by Kennicutt (1996) for a sample of spiral and irregular galaxies. The SFR histograms for the largest simulations (4 kpc) are somewhat narrower than the histograms constructed by Kennicutt for late type spirals or irregulars; however Kennicutt quotes a factor of two error in determining the birthrate. The long tail at large $b$ found for small-size simulations is qualitatively similar to the result found by Kennicutt for dIrr galaxies, which are in general smaller than spiral galaxies. The present interpretation, in which the larger-b tail is due to increasing burstiness for smaller galaxies, is in contrast to some previous interpretations which assume a SFR history that smoothly increases with time for late-type galaxies.

The widths of these distributions are approximately a factor of two greater than Poission statistics would predict ($\delta N = N^{1/2}$). This disparity is not surprising since Poisson statistics assume independent events and spatio-temporal correlations of the clusters are clearly present. In fact, given the large coherent areas of star formation present in the simulations, it is perhaps surprising that the fluctuations are not even larger.

### 4.1.2 Variation of star formation threshold

As the star formation threshold $N_{\rm sf}$ increases, the star formation rate decreases, since shells must grow to larger column densities before forming stars, while the average number of massive stars per cluster increases due to the assumed constant star formation efficiency and the larger gas mass available. The decrease in the star formation rate with $N_{\rm sf}$ causes fluctuations in the star formation history to grow as $\delta N_{\rm cl} \approx 2 N_{\rm cl}^{1/2}$. Figure 10 shows the star formation history along with the gas velocity dispersion for three values of $N_{\rm sf}$, all for a domain size $L = 2$ kpc. For the larger values of $N_{\rm sf}$ (in which the star formation criterion is more stringent making star formation more difficult), the star formation history becomes bursty.

Figure 11 shows the star formation histograms for four models like those presented in Fig. 9. These results suggest that, within the context of the present models, the observed distribution of present-to-average star formation rate parameter b is sensitive to the shell column density threshold for star formation, besides the size effect discussed earlier. Thus the broad b-distributions found in dIrr galaxies could be due to a larger column density threshold, for example because of lower metal abundance. Unfortunately, this suggestion cannot be tested on a galaxy-by-galaxy basis because the b-distributions of the simulations refer to single models sampled at different times. A given small or low-metallicity galaxy could have a small or large value of b, depending on when it is observed.

### 4.1.3 Time delay

As discussed in sec. 3.2.4, increasing the time delay between the onset of shell instability and the onset of the cluster wind, $\tau_d$, relative to the duration of the wind ($\tau_w = 10^7$ yr), produces star formation structure which is spatially more organized into coherent "superclusters," as seen in Fig. 6. This enhanced spatial coherence is expected to translate into greater temporal coherence of the global star formation rate evolution. Fig. 12 shows this effect. The global SFR is shown as a function of time for three simulations which all have the standard parameters except that the time delay $\tau_d$ is increased from 0.5 to 2.5 to 10 Myr. Although some increased temporal coherence can be seen in going from $\tau_d = 0.5$ to 2.5 Myr, a large effect is seen at $\tau_d = 10$ Myr, where the time delay equals the duration of the momentum input. Although the most evident fluctuations are seen to occur with timescales $\approx \tau_d$, one also sees correlated behavior over all timescales, similar to "brown noise." The SFR does not appear statistically stationary in the mean, although the departures from stationarity are marginal. As remarked earlier, larger values of the time delay are probably most appropriate if the cluster winds are driven by collective supernovae.

It is typical of nonlinear one-zone systems that oscillatory and even chaotic time dependence results when the delay time exceeds some other characteristic timescale of the system. Scalo & Struck-Marcell (1987) found, for a one-zone closed fluid model of star formation, that limit cycles (oscillations) in the system variables occurred when the time delay approached the characteristic timescale, which was the cloud collision timescale in their model. They also found chaotic (deterministic but unpredictable) behavior at still larger time delays. Unfortunately we have not yet explored models with $\tau_d > \tau_w$. Nevertheless, our simulations do illustrate that, even when spatial degrees of freedom are included, a large time delay between onset of instability and onset of energy injection may introduce substantial spatial and temporal coherence into the system, at least for the global scale $L = 2$ kpc investigated here. Whether or not larger time delays will result in more "bursty" behavior is a question left open for future work. However we point out that the relative fluctuations in SFR for the $\tau_d = 10$ Myr model are not appreciably larger than for the case with small $\tau_d$, although the average SFR is larger, as explained in sec. 3.2.4.

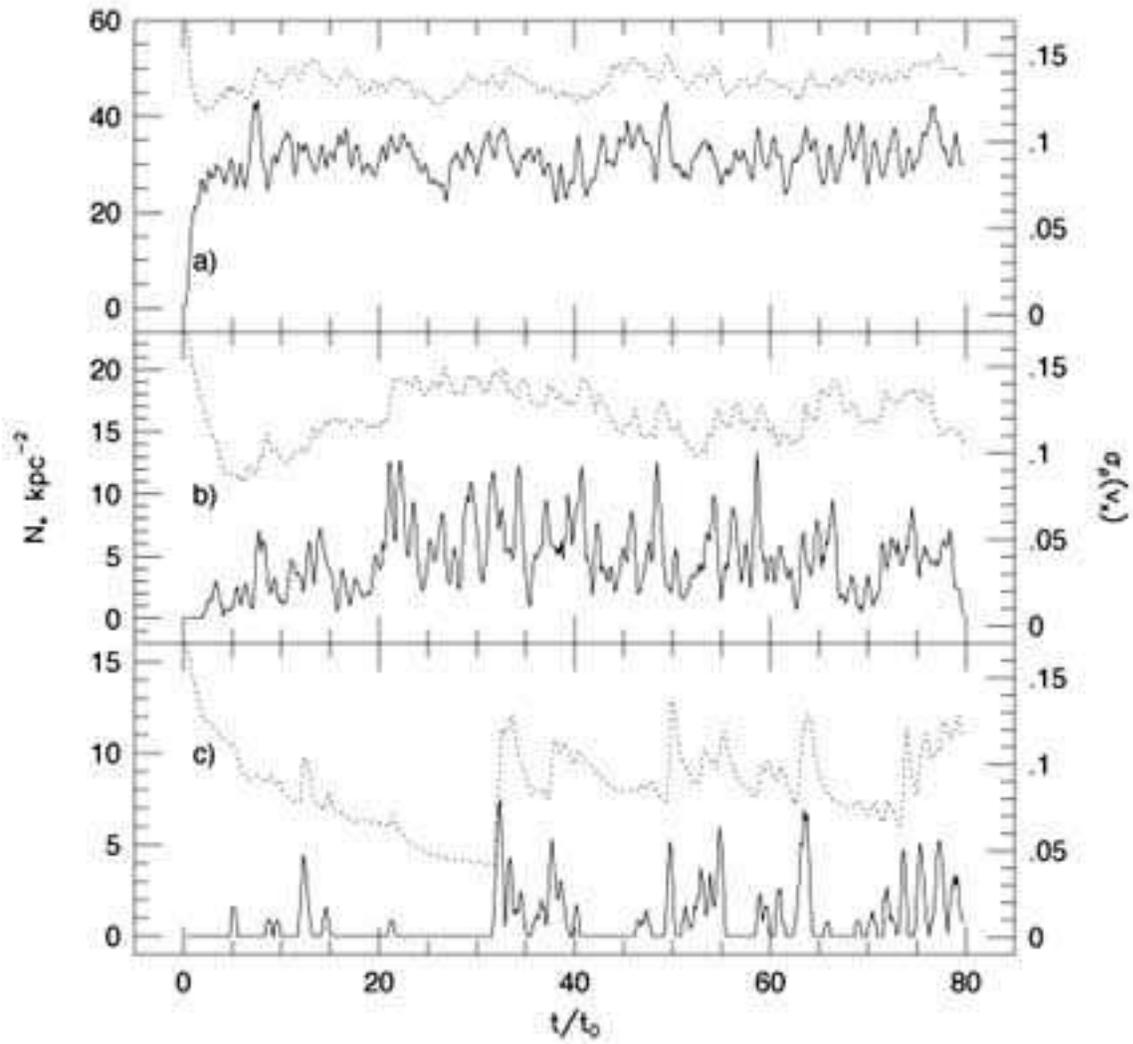

Figure 10. Evolution of global statistics for runs C with L = 2 kpc over 2 Gyr. Solid and dotted lines represent the number of massive stars per kpc$^2$ (left-hand axis) and the gas velocity dispersion (right-hand axis), rspectively. (a) $\tau_{sh} = 0.5$ km s$^{-1}$ ($N_{sh,21} = 3$), (b) $\tau_{sh} = 1.5$ km s$^{-1}$ ($N_{sh,21} = 1$), (c) $c_{sh} = 3$ km s$^{-1}$ ($N_{sh,21} = 0.5$).

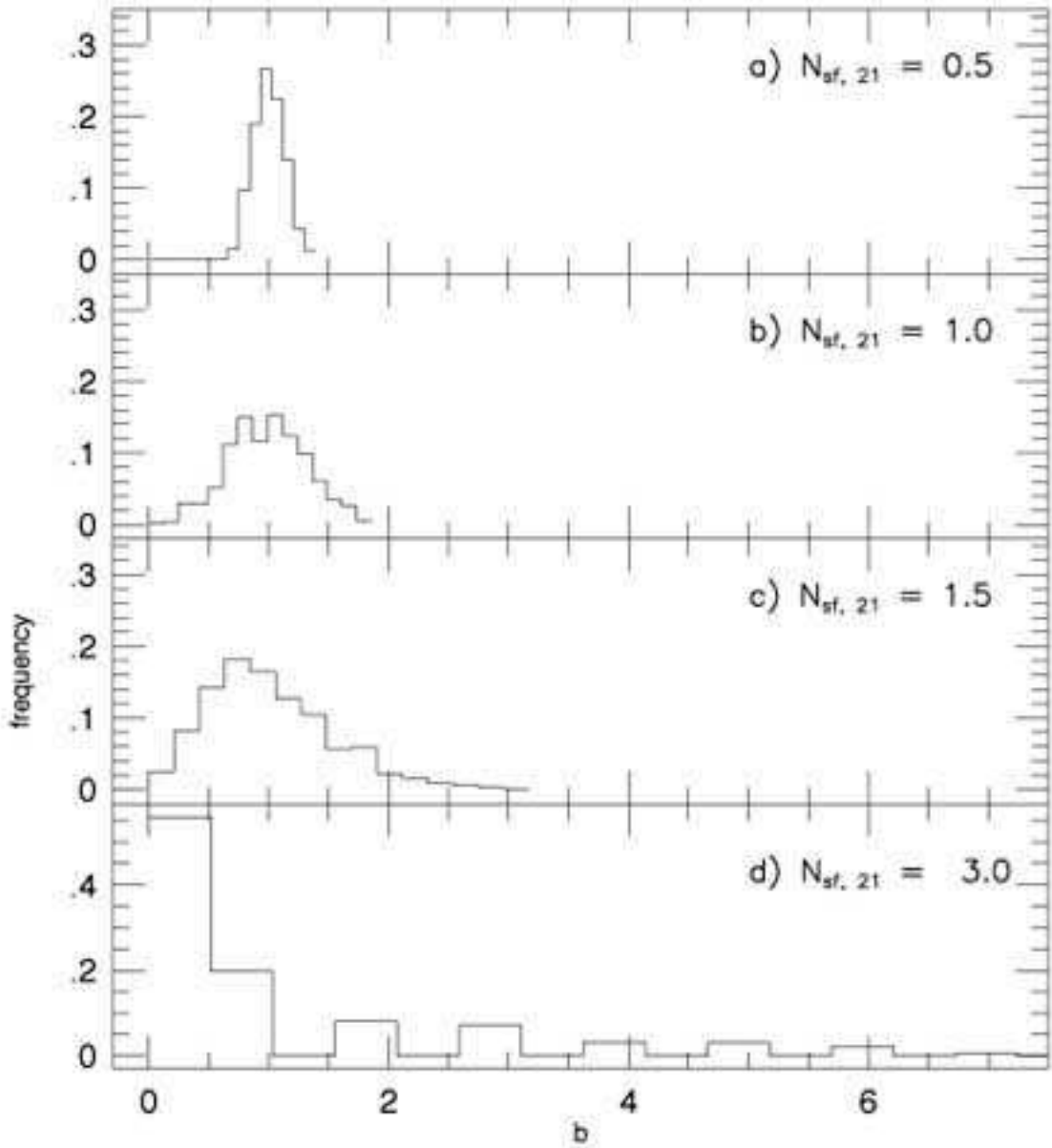

Figure 11. Histograms of present-to-=past average SFR ratio for four models with different adopted column density threshold for star formation. The top two and bottom panels correspond to the three time histories shown in Fig. 10.

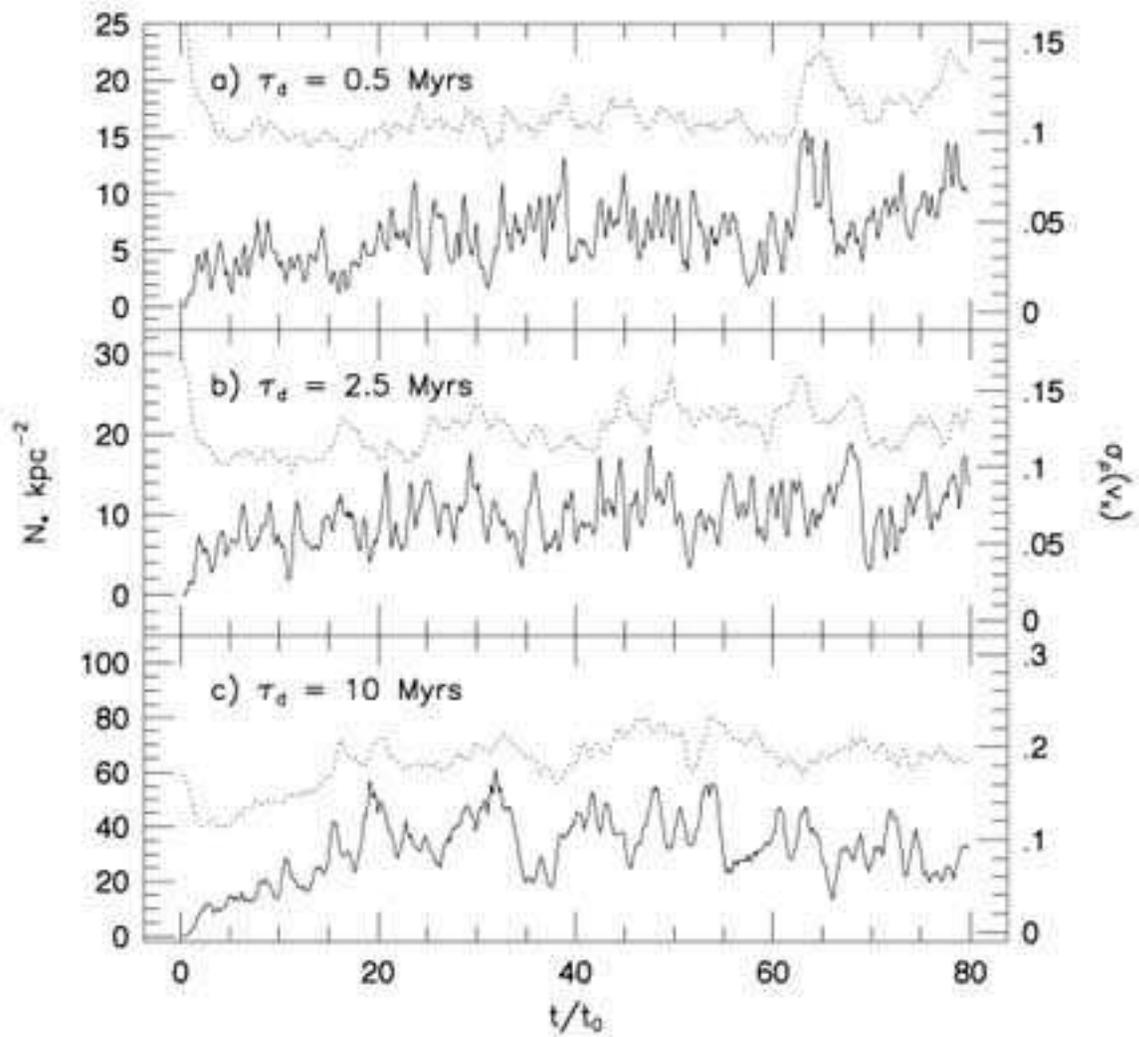

Figure 12. Time dependence of the global SFR (solid lines) and gas velocity dispersion (dotted lines) for simulations with time delay 0.5 Myr (a), 2.5 Myr (b) and 10 Myr (c).

### 4.1.4 Variation of average SFR with parameters

The variation of the time-averaged SFR (per kpc$^2$) with model parameters is of obvious interest. Although we ran a large number of models, we could not explore the complete parameter space, and in most cases only varied a single parameter with respect to the standard model. Still, some interesting results can be seen from inspection of Table 1.

First, for series A in which only the domain size was varied, the SFR (and most of the other global quantities listed in Table 1) are insensitive to the domain size between the $128^2$ (1 kpc$^2$) and $512^2$ (4 kpc$^2$) runs. However for smaller sizes the SFR decreases. Notice that the "burstiness" parameter $\sigma(b)/b$ increases continuously with decreasing size, showing that the temporal irregularity of the SFR is truly a function of size, not resolution.

The simulations of series C vary the shell internal velocity dispersion, or equivalently, the critical shell column density for star formation. Increasing $c_{sh}$ corresponds to decreasing $N_{sh,21}$, so star formation takes place more easily. However the SFR varies roughly quadratically with the shell velocity dispersion for runs Cg through Cq, although the dependence is somewhat steeper for runs Ca through Cf. This suggests that the SFR is controlled by binary interactions between shells, which is plausible since, as discussed earlier, most of the gas accumulated by a given shell consists of other shells driven by different star formation sites.

The dependence of the SFR on the time delay in series D follows a $\tau_d^{0.5}$ relation. Although we explained earlier why the SFR should increase with time delay, we can offer no physical explanation for the $\tau_d^{0.5}$ dependence. We note that if $\tau_d$ is inversely proportional to the square root of the density, as should occur for gravitational instability, this gives a contribution to the SFR dependence $\sim \rho^{-1/4}$.

The dependence of the SFR on the assumed average energy input per massive star (series E) is roughly linear, with SFR $\sim E^{0.8}$ from Table 1.

In addition to the models discussed so far, we also ran one series of five $256^2$ models in which the mean gas density was varied between values of 0.33 and 2 (the standard value used in all other simulations was 1.) All other parameters were the same as in the standard model except that the time delay was zero. These models are referred to as series F in Table 1. For the given internal shell velocity dispersion, and hence critical shell column density, the obvious main effect of increasing the average gas density is to make it easier for filaments to accumulate enough gas to drive them over the star formation threshold, increasing the star formation rate. Therefore this series should be similar to series C. In fact we do find that the SFRs of series F vary approximately quadratically with the mean density, again implicating shell interactions as the dominant process controlling star formation. However the models of series F had zero time delay. From the results of series D discussed above, we expect the time delay to introduce an additional factor $\rho^{-1/4}$ to the SFR if the time delay varies as $\tau_d^{-1/4}$. Thus the density dependence should be somewhat shallower than $\rho^2$.

However there are probably "hidden variables" in the SFR. Scalo & Chappell (1999) present a simple model for the velocity dispersion scaling relation which indicates that the SFR should depend on, besides the parameters discussed above, the overall velocity dispersion of the gas c and the average column density through filaments $\mu_{cl}$. The relation derived, which agrees remarkably well with the present simulations, is

$$N_* \sim c^3 \rho^{2/3}/(E\mu_{cl}) \qquad (16)$$

Thus the density dependence of the SFR cannot be determined observationally without explicit consideration of c and $\mu_{cl}$. The latter quantity would be especially difficult to determine observationally. This illustrates the danger in ascribing any physical significance to an observed SFR-density correlation.

### 4.2 Gas velocity PDFs

Figure 13 shows the probability density functions (pdfs), estimated as histograms, for three models from series C in which the star formation threshold is varied. The models are the same as for the density fields shown in Fig. 5. The threshold is smallest (SFR largest) for the top two panels, while the threshold is largest (SFR smallest) for the bottom two panels. The left-hand panels (a, c, e) include the entire density field, while the right-hand panels (b, d, f) include only those grid points that were identified as part of a filament according to the filament-finding algorithm. The open and closed circles represent the straight and mass-weighted velocity fields, respectively. The histograms are presented as log-linear plots, so an exponential (Gaussian) pdf would appear as a straight line (parabola). The histograms for the full field (left-hand panels) are based on five velocity fields sampled every $2 \times 10^8$ yr for the last Gyr of the simulations, so each composite histogram of these $256^2$ runs contains $3.3 \times 10^5$ points. They exhibit a low-velocity core and a broader high-velocity component. The core appears nearly exponential, although a Gaussian is not excluded for the smallest velocities, while the broad tails have an even slower, possibly power-law fall-off. The broad tails reflect the velocity distribution of the winds themselves, i.e. regions interior to the shells.

Even more striking are the velocity histograms restricted to the filament ridges (right-hand panels). Since on average only a few percent of the lattice sites fall on a filament ridge, we combined filaments from 50 time steps equally spaced over the last half of the simulations (over which the global statistics appear statistically stationary;

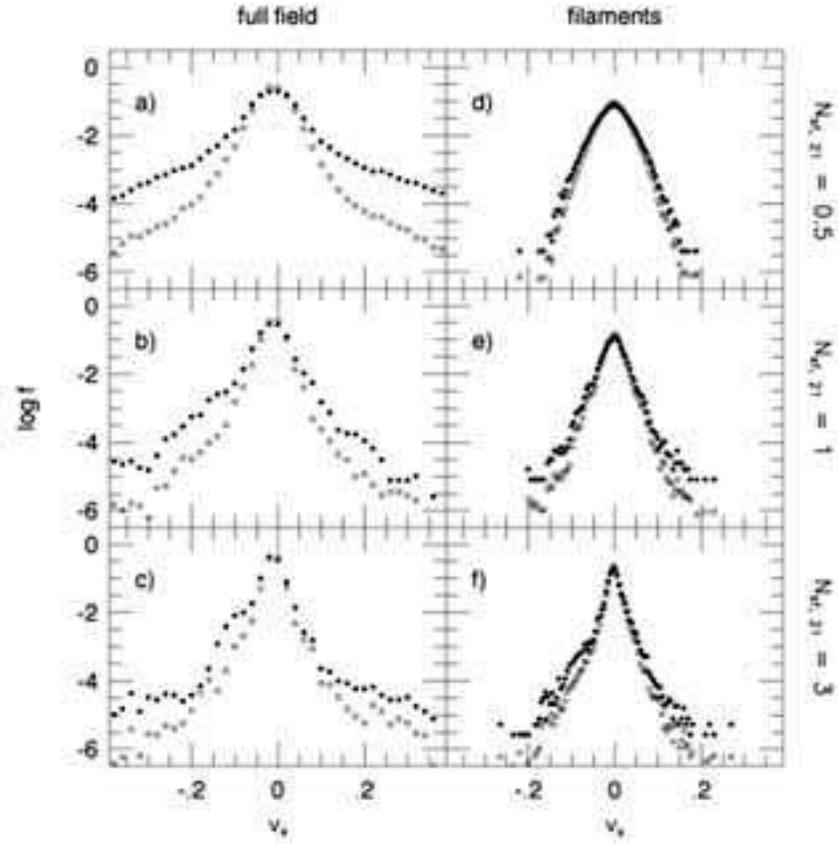

Figure 13. Velocity field histograms for series C with $L = 2$ kpc. The shell velocity dispersions are $c_{sh} = 0.5$ km s$^{-1}$ ($N_{sf,21} = 3$), 1.5 km s$^{-1}$ ($N_{sf,21} = 1$), and 3.0 km s$^{-1}$ ($N_{sf,21} = 0.5$) from top to bottom. The left columns are composite histograms of the x component of the total velocity field over five times for each simulation. The closed and open circles correspond to straight and mass-weighted velocities. In panels b, d, and f only those points found to lie along a filament ridge are included in the histograms.

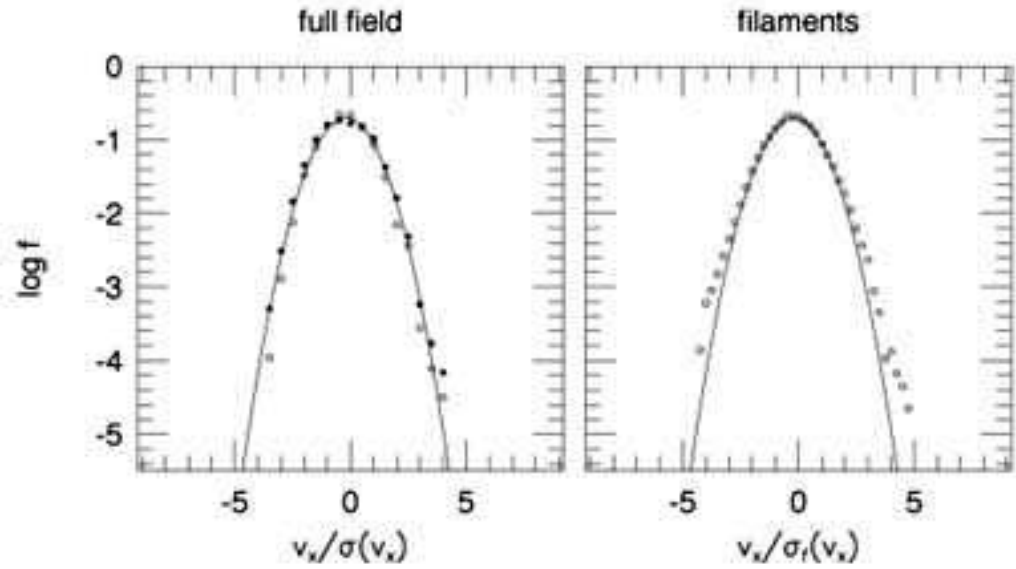

Figure 14. Velocity histograms for $512^2$ decay simulations (no star formation) with a scale-free initial energy spectrum $E(k) \propto k^1$. *Left*: histogram of x-component of the total velocity field. *Right*: Only points found to lie along filament ridges are included. The parabolic reference curve corresponds to a Gaussian distribution. Comparison with Fig. 13 suggests that the tail excesses for the gas in filaments is not due to star formation activity.

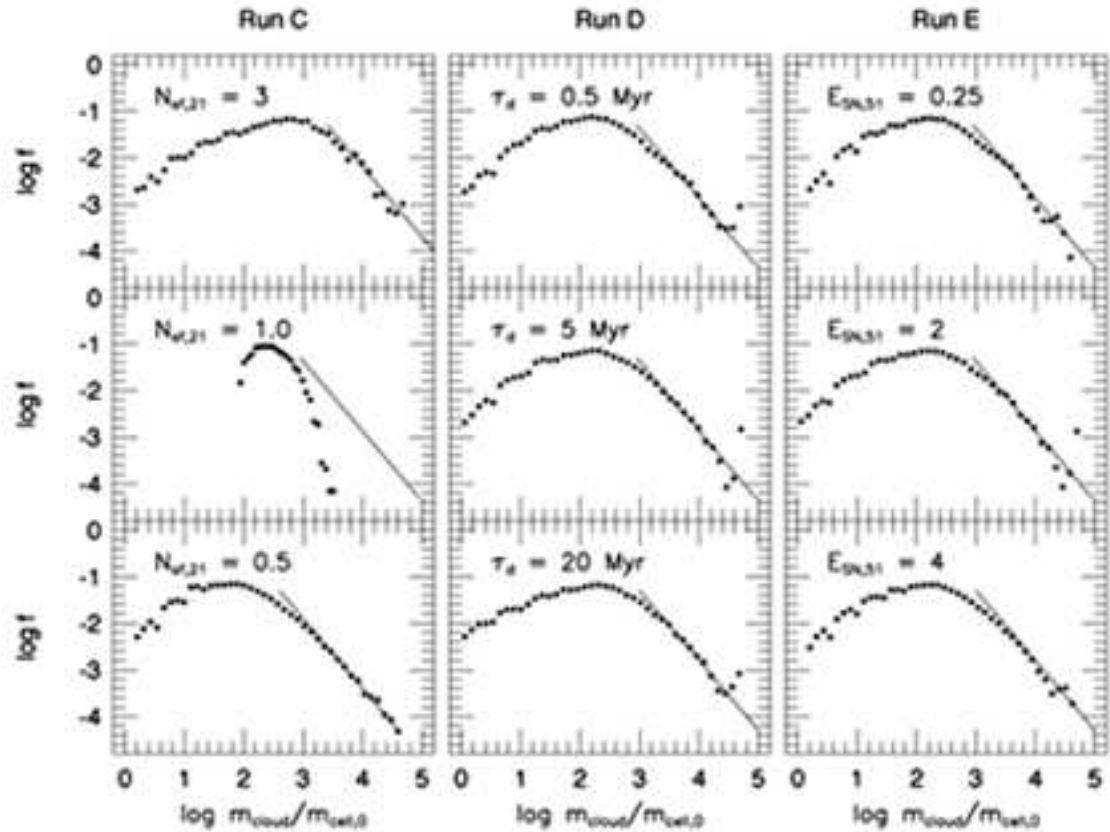

Figure 15. Cloud mass functions (number of clouds per unit logarithmic mass interval) for three parameter choices in each of the model series' C (varying star formation threshold), D (varying time delay), and E (varying assumed energy input per supernova event). Each mass function was calculated by the structure tree method applied to 100 times throughout the last Gyr of each simulation. Solid lines at large masses have power law index –1.5.

see Fig. 10) to construct the histograms, resulting in about $10^5$ points per histogram. These histograms are very nearly exponential except for the possibly power-law flattening in the far tails, which is most apparent in the lowest SFR run (Fig. 13f). In no case is there evidence for a Gaussian component.

It might be thought that the exponential pdfs are the result of the action of stellar winds in the simulations, which continually produce high-velocity filaments. In fact this is not the case. Instead, we propose that the exponentials are a consequence of momentum-conserving (but not kinetic energy-conserving) filament interactions. In support of this interpretation, we show in Fig. 14 the velocity pdf for a typical pure decay simulation (no star formation). The full field, shown on the left, is nearly Gaussian, demonstrating that the broad tails of the full velocity field pdfs shown in Fig. 13 are a direct result of the star formation-driven winds. However the pdf of the material in filament ridges (right) shows a clear excess over this Gaussian for intermediate and large velocities. (Note that the units of the velocity axis are different in Fig. 14 than in Fig. 13.) The excess at large velocities appears exponential, even though no wind-driven filaments are present. The reason the exponential tails are not as prominent as in Fig. 13 is due to the fact that, in the decay simulations, the number of filaments and their velocities continually decrease with time, so the time between filament interactions becomes large, limiting the number of filament interaction which can have occurred over the duration of the simulations. In contrast, the simulations which include star-formation-driven winds (Fig. 13) continually supply new high-velocity filaments, so the number of filament interactions which have occurred at a given (late) time is much greater. We therefore suggest that the exponential velocity pdfs of the filaments is a result of multiple inelastic shell interactions.

Analytical models for the exponential and power law for tails of the velocity pdfs are given in a separate paper (Scalo et al. 2000), and we are mainly concerned with presenting the simulation results here. The analytical explanation for the exponential portion of the pdfs centers on the fact that the physical system is one in which mass and momentum, but not energy, are the global conserved quantities. One way to understand this is by examining the statistics of a Gibbs ensemble in which there is no quadratic invariant (kinetic energy). Another related, but less abstract, explanation involves the statistics of momenta of interacting filaments which have undergone a number of collisions that is Poisson distributed. The power-law far tails for the low-SFR runs can be understood in terms of the predicted velocity pdfs of filaments or shells which have not undergone many interactions (see Silk 1995). At the other extreme, for a very large number of filament interactions, the central limit theorem or a Langevin equation approach suggests a Gaussian pdf. Details are given in Scalo et al. (1998).

Independent of theoretical interpretation, our simulation results are broadly consistent with the observational centroid velocity pdfs presented by Miesch & Scalo (1995) and Miesch, Scalo & Bally (1999) for regions of active internal star formation, and with the results of optical absorption line and H I emission and absorption line centroid velocities presented in Miesch et al. (1999), both of which suggest exponential pdfs for most regions. An exponential distribution of H I fluctuation velocities out of the plane of the LMC was found by Kim et al. (1998). Our finding of possible power-law contributions in the far tails of the pdf, which we attribute to stellar wind sources, is consistent with the analysis of the high-velocity optical line pdf given by Siluk & Silk (1974).

### 4.3 Cloud mass spectra

We constructed "cloud" mass spectra by employing an algorithm similar to that used in computing structure trees (Houlahan and Scalo 1992). The method is based on thresholding the density field at a number of intensity levels (logarithmically spaced in the present analysis and given by $\rho_{\text{thresh}}^i = 2^{i/2}\rho_0$ where $i$ is a non-negative integer) to form a contour image. Each closed contour is linked-up to its "parent" (defined as the next lower contour which contains the original contour of interest) if the parent has no other "children" or if the area of the original contour is larger than all of its siblings at that level by a predetermined factor (of order 2). If the region is *not* linked up, it is defined to be a cloud; otherwise it is assumed to just be part of a larger cloud defined at a lower threshold.

This technique provides a sophisticated yet simple method of identifying clouds which avoids the problems encountered by counting all closed contours over either a single threshold ratio or many thresholds (e.g. Stutzki et al. 1991, Williams et al. 1994). The mass spectra constructed from closed density contours at a single, arbitrary contour level not only depend on the contour level chosen, but disagree with other techniques. In addition, methods which use many contour levels introduce a bias in which clouds are over-counted if they appear at multiple threshold values. With the present technique, a cloud that contains no substructure is counted only once in the mass spectrum, independent of the number of density contour levels chosen. The method is also not "fooled" by small density fluctuations that show up as small closed contours superimposed on larger cloud contours. Such cases may either be rejected as noise or counted as condensations in the larger cloud; in either case, the identity of the larger cloud is not affected. In computing the mass spectra for the simulations, we rejected all closed contours with areas less than five simulation cells from the mass spectra.

Application of this procedure to the simulation gas distributions produces a few hundred clouds per time step. Thus, to build up a statistical sample we construct a mass spectrum based on the density field at 100 times throughout the last Gyr of the simulation, giving on average $10^4$ points. Figure 15 shows examples of the resulting composite cloud mass spectra for three models for three different series. (The mass spectra shown in the figure represent the number of clouds per unit logarithmic mass interval $dN(m_{\text{cl}})/d\log m_{\text{cl}}$. Conversion to linear mass interval $dN(m_{\text{cl}})/dg m_{\text{cl}}$ steepens the power-law slope by a factor of $m_{\text{cl}}$.) At large masses we find that all the differential mass spectra scale approximately as $dN(m_{\text{cl}})/dm_{\text{cl}} \propto m_{\text{cl}}^{-2.5}$ independent of the model parameters.

Comparison with observationally determined mass spectra is not straightforward since the clouds derived from the simulation gas distribution are two-dimensional and estimation of three-dimensional masses depends on assumptions about the scaling between the vertical extent of the clouds out of the simulation plane with their 2-D properties. An instructive way to see that the 3D mass spectrum corresponding to a given 2D mass spectrum should be flatter is to assume that, given the same physical processes, the characteristic sizes of structures should be the same in 3D as in 2D. More specifically, we assume that the pdf of structure characteristic sizes, p(r), would be the same in 3D as in 2D, and assume it is a power law given by p(r)$\sim r^\alpha$, with $\alpha < 0$. We then compare the differential mass spectrum $f(m) = p[r(m)]dr/dm$ in the 2D and 3D cases.

Inspection of simulation density images at various thresholds shows that most of the entities identified as "clouds" by the structure tree method are filamentary. We assume that in 3D these structures would be shells. Then, in 2D, the mass of a cloud is $m \sim r\rho$ and the mass spectrum corresponding to a given size spectrum power law index $\alpha$ (assuming $\rho = $ constant) is $f(m) \sim m^\alpha$. But in 3D we have, for shells, $m \sim r^2\rho$, and obtain $f(m) \sim m^{(\alpha-1)/2}$. Since $\alpha$ is negative, this shows that the mass spectrum in 3D will be flatter than the 2D simulation mass spectra by an amount $-(\alpha+1)/2$ in the mass spectrum power law index. The flattening varies from $1/2$ for $\alpha = -2$ to $1$ for $\alpha = -3$. For the particular mass spectrum found in the 2D simulations, $f(m) \sim m^{-2.5}$, the flattening is $3/4$, so the 3D mass spectrum should be $m^{-7/4}$. This result is equivalent to the result found by simply assuming that the filaments are portions of shells which, in 3D, would have vertical extent comparable to the 2D filament length (an assumption which would not be justified for structures of extent larger than the vertical scale height). If the index of the 2D mass spectrum is $\gamma$, then the index in the 3D case will be $(\gamma - 1)/2$, which is a flattening (as long as $\gamma < -1$).

Additional flattening in 3D is found if we assume that the densities are related to size as $\rho \sim r^{-p}$. In that case it can be shown that the flattening in going from 2D to 3D is $-(\alpha+1)/(p^2 - 2p + 2)$. For any $p < 2$ this flattening is in excess of the flattening found above for the constant density case. For example, if $p = 1/2$, the 2D case with $f(m) \sim m^{-2.5}$ gives $\alpha = -7/4$, and then the 3D mass spectrum should be $m^{-3/2}$.

In reality there is not a one-to-one correspondence between density and size in the simulations, although the correlation does exist. In the other extreme that density and size are independent random variables, one could in principle use the density pdf and the assumed size pdf to calculate the pdf of masses $m \sim r\rho$ (2D) and $m \sim r^2\rho$ (3D) using procedures for calculating the pdf of products of random variables. However this procedure would require the inversion of Mellin transforms which do not exist in closed form, and so we do not pursue this approach further.

To summarize, the present 2D simulations predict differential mass spectra that are power laws at large masses, with index –2.5, independent of the simulation parameters. Arguments have been presented that suggest that, if the simulations were 3D, the mass spectra would possess large-mass mass spectra that are flatter by roughly unity in the power law index.

A large body of observational results for molecular clouds find a mass spectrum slope of $-1.3$ to $-1.9$ over a large range of masses (see Solomon et al. 1979, Casoli et al. 1984, Dame et al. 1986, Blitz 1993, Brand & Wouterloot 1995, Kramer et al. 1998, Heyer & Tereby 1998, and references therein), although there are exceptions (e.g. Onishi et al. 1996 for Taurus cores). Mass spectra with a power-law exponent of $-1.5$ have been derived through non-dynamical equilibrium coagulation models which assume the cloud velocity dispersions and cross sections are either mass-independent or have some simple form (e.g. Field & Saslaw 1965, Kwan 1979). However the theoretical results do depend on the above assumptions (see Silk & Takahashi 1979, Elmegreen 1989). For example, Elmegreen (1989) argued that if the cloud column densities are assumed constant in deriving the cross section, coagulation models should predict a steeper cloud mass spectrum, with $n(m_{cl}) \propto m_{cl}^{-2}$. This result was used to suggest that cloud coagulation models are inconsistent with observations. (For discussions of how selection effects could be responsible for an apparently constant column density, see Kegel 1989 and Scalo 1990). N-cloud simulations of coalescence models (e.g. Pumphrey & Scalo 1983, Nozakura 1990) can directly treat the cloud velocity distribution, unlike the kinetic equation approach, but still depend on the assumed coalescence cross section and the assumption of discrete independent clouds (rather than an advecting fluid). In contrast, the present simulations are entirely fluid-dynamical and involve no assumptions about cross sections or density-size or velocity-size relations, although they are limited to two dimensions. Perhaps most importantly in the present simulations advection can stretch or compress clouds into filaments, and spatial correlations between structures of various sizes are self-consistently accounted for.

Our results show that the 2D mass spectra at large masses are approximately power laws whose slopes are insensitive to parameters, and we argue that the corresponding 3D mass spectra are roughly consistent with observations. However we want to point out that our argument concerning the difference between 2D and 3D mass spectra is not really self-consistent; for example, if filaments had a vertical extent comparable to their length, their masses and momenta would evolve differently, perhaps altering some essential feature of the coagulation process. Clearly a 3D simulation with a careful cloud-counting algorithm will be required to estimate the 3D mass spectrum.

## 5 MULTIPERSPECTIVAL INTERPRETATION

One of the most interesting features of the present work is that it can be regarded as a simulation of several conceptually different models for the evolution of structures and star formation in galaxies, since its aspects include collisional coalescence of structures, gravitational instability in shells, propagating star formation, self-regulation, the dynamics of a field of advection-created discontinuities or shocks, and wind-driven (entirely) compressible turbulence. None of these aspects are more, or less, important than the others; they are different, incomplete, ways of viewing a complex phenomenon, whose complexity has already been severely truncated in the simulations (omission of pressure, magnetic fields, etc.). We feel that this "multiperspectival" interpretation is an important result, for it suggests that a particular conceptual model, while perhaps useful in describing some aspect(s) of the observed phenomena, cannot be expected to provide a complete description, and will therefore be "invalidated," sooner or later, by observations that capture aspects excluded by the particular model. Similarly, when discussing conceptual models for galactic gas dynamics, a model may not be more, or less, valid than another model, if the two models are not meant to explain the same aspect. For these reasons we briefly summarize the simulations and compare with a number of previously-investigated conceptual models.

In the present simulations, star formation occurs on a wide range of scales. First, as the wind-blown shell from a star formation event expands through the local gas complex it may trigger the birth of more stars by sweeping up the surrounding gas, pushing the shell column density over the critical value for star formation. This mode of star formation corresponds to the scenario pictured by many propagating star formation theories which consider a single isolated shell expanding into a uniform and quiescent background medium (e.g. Elmegreen & Lada 1977, McCray & Kafatos 1987, Comeron & Torra 1994, Whitworth et al. 1994). However, since the local gas complexes in the present simulations (and in the observed ISM) are often irregular or filamentary, the distribution of the triggered star formation events are also highly irregular in appearance, often reflecting the underlying gas distribution. This local propagation mechanism is found to occur in a significant fraction of star formation events.

Second, star formation also proceeds through the interaction of shells. When the interaction of relatively young shells leads to star formation, the propagation sequence is clear. However, star formation events involving the interaction of older shells which have undergone repeated mergings are the product of many initial "triggering" shells. In this case, star formation may also be considered as being "spontaneous," although the physical process involved is the same as in the more local interactions. The formation of gas concentrations, and thus stars, through this mechanism may be viewed as either taking the zero pressure limit of the hydrodynamic hydrodynamic equations or including fluid advection into a coagulation model.

The star formation rate, controlled in part by the star formation threshold, is found to control the structure and dynamics of the gas. At low SFRs, the gas has time to coalesce into dense filaments leaving large empty regions of low-density gas. In addition, young shells rarely interact and the probability distribution of the velocity field has the power-law form of a set of non-interacting shells. Also, we find that in this case the energy spectrum is found to have the $E(k) \propto k^{-2}$ form predicted for a field of shocks. As the star formation rate increases the filaments become increasingly shredded and shell interactions become increasingly important. These interactions cause the the velocity pdf to become nearly Gaussian at small velocities while the tail takes on an exponential form. In addition, the energy spectrum slightly flattens as the correlation length of the shredded shells decreases.

We find that these mechanisms lead to the emergence of coherent and dynamic regions of star formation on spatial scales comparable to associations, complexes, and supercomplexes of OB stars. The loci of these star formation regions move throughout the simulation, following complex and "unpredictable" trajectories. The spatial clustering distribution of the young stars is scale-free, as evidenced by the power law correlation functions presented in Scalo & Chappell (2000).

These simulations may be viewed from several perspectives.

First, the present work may be thought of as a simulation closely related to the stellar-driven turbulence models of Norman & Silk (1980) and Norman & Ferrara (1996), who invoke severe approximations to model turbulence with analytic techniques. While Norman & Silk (1980) picture the shells as expanding into a turbulent background formed by a network of old momentum-conserving shell fragments, very similar to the present simulations, they model the turbulence through a cloud coalescence equation in which the basic nonlinearities governing turbulence are ignored. Norman and Ferrara (1996), on the other hand, use a transfer function applicable to *incompressible* turbulence to model the evolution of the energy spectrum. On the contrary, most of the interactions in a galactic ISM should be highly compressible. Thus, results from the present model which explicitly model the effects of fluid advection, compressibility, and small-scale forcing due to stars are more general (not being restricted to the evolution of the correlation function) and correspond more closely to the supersonic conditions expected in the ISM. However it should be emphasized that the results of the present simulations do correspond rather remarkably to the conceptual model of Norman & Silk (1980).

Second, the present reaction-advection simulations may be viewed as a model of propagating star formation analogous to either the cellular automata model developed by Gerola & Seiden (1982, see sec. 1 for later developments) or the reaction-diffusion models cited in sec. 1. In our simulations, however, the distinction between "triggered" and "spontaneous" star formation is not a sharp one. The primary mode of star formation in our simulations is the collision between dense shells or clumps. In some cases, a shell sweeps up and compresses enough gas to satisfy the star formation condition while the shell is still young and roughly circular. Such an event satisfies the traditional picture of propagating star formation in which a single shell expanding into a uniform ambient medium leads to the formation of more stars. However, the interaction of shells, after they have been distorted by the ambient velocity field and possibly merged with other shells, also leads to star formation through the coalescence mechanism. In this case many initial stars spread over a large region of space could be viewed as being equally responsible for the "triggering."

Third, since the fluid is highly inelastic, the simulation may be viewed as a coalescence model which explicitly includes the effects of fluid advection and the velocity field of the gas. A basic question is whether the inclusion of fluid advection affects the form of the cloud mass spectra predicted by coalescence models (e.g. Field & Saslaw 1965, Kwan 1979, and Elmegreen 1989). Our models do not involve any assumptions concerning the form of the collision cross section or the "cloud" velocity distribution, both of which are completely determined by the solutions of the reduced hydrodynamic equations. We find that the cloud mass spectrum at large masses is a power law with an index that is insensitive to parameters. However the precise form of the predicted 3D mass spectrum is uncertain.

Fourth, the global gas structure found in our simulations, especially in the case of no star formation (de-cay runs), is remarkably similar to that found in two-dimensional simulations of granular fluids by Goldhirsch & Zanetti (1993), when the assumed coefficient of restitution is small enough. The particles in the latter simulations are assumed to be disks of identical sizes which collide with geometric cross section and a given coefficient of normal restitution. The granular system, like our galactic gas model, conserves mass and momentum but not energy. Because there are a very large number of grains, a continuum description is possible (although with certain difficulties concerning the proper definition of thermal variables), resulting in a nonlinear advection term in the momentum equation. This similarity suggests that the overall spatial distribution is generic to inelastic systems that conserve mass and momentum, and that much of the behavior that we have found is fundamentally linked to the absence of a kinetic energy invariant. This idea will be used to understand the exponential velocity distribution function of filaments in a separate paper (Scalo et al. 1999, in preparation). The analogy with granular fluids also suggests a generic relation between granular fluids and mass-conserving Burgers turbulence (see below).

Fifth, we refer to this model as a "reaction-advection" model to contrast it to the reaction-diffusion systems mentioned in sec. I. We suggest that a reaction-advection model is more appropriate to the problem of propagating star formation than reaction-diffusion models because it embraces more fundamental aspects of the underlying system such as gas conservation laws and advection. Furthermore, besides omitting important phenomena (shocks due to advection), reaction-diffusion models have the danger of introducing spurious phenomena (e.g. Turing patterns that depend on the dominance of diffusion). In addition, we suggest that generic forms of this reaction-advection model may find applications to problems in other fields in which advection plays a significant role. As far as we know, this is the first study of this class of systems.

Closely related to the above view, this model can also be regarded as a reactive Burgers equation, since in the absence of stellar forcing the equations approximate Burgers' equation (which is simply the pressureless Navier-Stokes equation). A difference from the usual formulation of the Burgers problem is that we explicitly enforce mass conservation by solving the continuity equation as well as the (pressureless) momentum equation. It is interesting to note that Burgers originally studied this equation as a possible simplified model of Navier-Stokes turbulence, but it was later pointed out (Kraichman 1965) that it lacked a mechanism to destroy correlations. However, since the presence of the reaction term (star formation here) does provide this effect, we suggest that the reactive Burgers equation (with the inclusion of mass conservation) may be an interesting and relatively simple model of highly compressible turbulence. The study of "Burgers turbulence" is an extremely active field of research (see Bernard & Gawedzki 1998, Gotoh 1999, E & Vanden Eijnden 1999, Fogedby 1999, Ryan & Avellaneda 1999 and references given there).

## 6  SUMMARY

We explored the ways in which stellar feedback and fluid advection can shape the morphology of the gas density field, the structure of the gas velocity field, and the spatio-temporal organization of the stars themselves. The simulations follow the evolution of a system of interacting wind-driven shells (which obey global mass and momentum conservation laws and are subject to fluid advection). This allows us to investigate, in a self-consistent manner, the physics of wind-driven turbulence modeled by Norman & Silk (1980) and Norman & Ferrara (1996), propagating star formation models studied through the use of cellular automata by, for example, Seiden & Gerola (1982), Comins (1983), Jungwiert & Palous (1994), and Pedang & Lejeune (1996), and diffusion approximations by, for example, Shore (1983), Korchagin & Ryabtsev (1992), Neukirch & Feitzinger (1988), and Nozakura & Ikeuchi (1984, 1988). An important advantage of our simulations over these models is in our treatment of the "turbulence" through direct simulation of the nonlinear advection operator. One-zone models, by definition ignore spatial interactions altogether. Diffusion models include spatial couplings but in a linear and, we think, dangerously inappropriate manner. Norman & Silk (1980) model the shell-shell interactions through a coagulation equation which neglects, or at least highly abstracts, the nonlinear turbulent processes, while Norman & Ferrara (1996) use a transfer function (in Fourier space) for *incompressible* gas to model the effects of the nonlinear advection operator, and even then only at the level of the correlation function. Furthermore, most studies of propagating star formation ignore the gas dynamics altogether. In addition, our simulations, in which star formation acts as a "reactive" forcing term, can be compared to studies of Burgers' equation with external forcing. We of course do not claim that our simulations should serve as substitutes for simulations of the full hydrodynamic or MHD equations that include many physical processes we have excluded (e.g. Passot et al. 1995, who include self-gravity, the magnetic field, star formation feedback, heating and cooling, and rotation). However we do claim that intermediate levels of hydrodynamic modeling such as that of the present paper, can yield much insight from an explanatory point of view, and also represent a more computationally efficient and structurally simpler approach than simulating the full hydrodynamic or MHD equations. An interesting example of this approach is the use of one-dimensional magnetic Burgers equations (Yanese 1997) to model solar flares and coronal heating (Galtier & Pouquet 1998).

We now review some of the results and predictions from these simulations.

(i) The spatial distribution of the gas in all the simulations evolves into a complex network of interacting partial shells very similar to the conceptual model envisioned by Norman & Silk (1980) and to the observed morphology reported by Chu & Kennicutt (1994) for 30 Dor in the LMC, Kim et al. (1998) for H I in the entire LMC, Mizuno et al. (1995) for the local Taurus complex, and Bally et al. (1999) for the local Circinus molecular cloud. The term "churning" used by Bally et al. is usefully appropriate for the present simulations. This dynamical network morphology seems to be an inescapable consequence of combined effects of stellar outflows, nonlinear advection, and high degree of gas compressibility. Although the inclusion of gas pressure, magnetic fields, and other processes will modify the results (especially the thickness of the shell structures), it is hard to see how these processes could qualitatively alter the general morphology, which can be seen in the most physically detailed previous simulations (Passot et al. 1995).

(ii) Star formation organizes into coherent structures over a range of spatial scales. The locus of these structures move throughout the grid leaving behind spatial gradients in the stellar ages. The sizes of these coherent structures correspond to typical sizes of OB associations, complexes, and supercomplexes. The correlation functions for the stars are scale-free power laws (Scalo & Chappell 1999). While the gas distributions that give rise to these structures involve a coagulation process, they also require feedback by star formation and the action of fluid advection. Thus, simple cloud coagulation models which do not incorporate such effects would miss this organizational mode.

(iii) The global star formation rate per unit area is found to be independent of the simulation size for all but the smallest simulations studied. Fluctuations in the global SFR are found to grow in amplitude as the system size decreases. These fluctuations have amplitudes roughly twice that predicted by Poisson statistics for all simulations studied and develop temporal correlations for simulations with spatial dimensions less than about a kiloparsec. Similar results were obtained by Gerola et al. (1980) in their Stochastic Self-Propagating Star Formation (SSPSF) model. They interpret the large fluctuations, in which long lulls occur in the global SFR, as as a mechanism by which dwarf galaxies can develop bursts. The present simulations, which directly model shell evolution and interactions and do not require the ad hoc assumptions of triggering and spontaneous probabilities, support this view. However percolation, which controlled the small-galaxy size bursts in the SSPSF models, plays no role in the present simulations. We also find that besides the size effect, the broad distribution of present to past average SFRs in late-type galaxies can be accounted for if late-type galaxies have smaller metal abundances, corresponding to a larger critical shell column density for gravitational fragmentation.

(iv) The SFR varies with the density as $\rho^{1.7}$ if the time delay scales as $\rho^{-1/2}$. However we find that the SFR also depends on the gas velocity dispersion and on the average shell column density, so observed univariate correlations of SFR with density should be regarded with caution.

(v) The probability distributions of filament velocities are strikingly exponential. The possibly power-law tails found at low star formation rates may reflect the power-law distribution of a set of noninteracting shells. The origin of the exponential tails, however, appears to be more universal, since they are also found for the decay runs with no star formation, and we suggest that these tails are due to multiple dissipative interactions of shells. This range of forms of the velocity pdfs appears roughly consistent with the empirical results of Miesch et al. (1999) for local molecular clouds and Kim et al. (1998) for H I in the LMC.

(vi) In the absence of star formation, we find the probability distribution function (pdf) of filament velocities to have a Gaussian core with excess, possibly exponential, tails. This is a new result which, to our knowledge, has not been previously reported in the Burgers turbulence literature. The tail excesses are used to support our contention that the exponential filament velocity pdfs found for simulations that include star formation are not primarily due to the cluster winds but are a result of shell interactions driven by highly compressible fluid advection.

(vii) The power-law slope of the cloud mass spectrum at large masses is found to be insensitive to the parameters. The majority of the runs scale as $dN(m)/dm \propto m^{-2.5}$ in 2D. Comparison of this result with observations is limited due to the two-dimensionality of the simulations. However, we presented arguments that the 3D mass spectrum should be significantly flatter, probably consistent with existing observations.

The central role that shell interactions play in structuring star formation in these simulations suggests that their detailed study could lead to a better understanding of propagating star formation. While the stability and evolution of single shells has been investigated through analytic methods (Vishniac 1983, Ryu & Vishniac 1988, Vishniac 1994, Elmegreen 1994, 1999) and simulations (Mac Low & Norman 1993, Blondin & Marks 1996), very little work has considered shell interactions. Stevens et al. (1992) considered a pair of interacting winds from a binary system, but this work is not directly applicable to the interaction of large, slower-moving shells, although it does illustrate the complexity of phenomena which may be expected. A study of the interaction of two shells generated by galactic explosions in a cosmological context (Weinberg et al. 1989) has been presented by Yoshioka & Ikeuchi (1990; see also Yoshioka & Icheuchi 1989). Although the scales of the relevant variables (size, temperature, etc.) are much different than in the present simulations, the non-dimensional situation is not far from the conditions used here, or which might occur within molecular clouds. This result suggests how the present simulations may be of relevance, in a generic sense, to phenomena that occur in a variety of astrophysical environments from the cosmological to the protostellar. Analytic approaches to the interacting wind problem have also been developed (see Canto, Raga & Wilkin 1996, and references therein). We suggest that further detailed hydrodynamic simulations of the interactions between wind-blown shells in a variety of physical contexts is needed. In addition, the detailed effects of winds from young "triggered" stars embedded in an expanding shell on the shell itself would be of interest.

Finally, we emphasize that our approach has focused on an intermediate-level, "reduced" description of the hydrodynamics, in which several potentially-important processes have either been omitted (pressure, magnetic fields, heating and cooling, rotation) or incorporated as externally-imposed "rules" (gravitational instability). Our conclusions therefore need to be checked by three-dimensional simulations that include these physical effects.

This work was supported by NASA Grant NAG5-3107. We thank Anthony Whitworth, Enrique Vazquez-Semadeni, and an anonymous referee for helpful comments.

# APPENDIX A: TWO-DIMENSIONAL DONOR-CELL ADVECTION SCHEME

The 2-D donor-cell advection scheme adopted for the present simulations was based on the methods discussed by Van Leer (1977) and Yanagita & Kaneko (1995). While the donor-cell scheme is an Eulerian technique with a fixed grid, it is convenient to imagine that on each time step, the $i^{th}$ cell is transported a distance $\mathbf{v}^i \Delta t$, where $\mathbf{v}^i$ is the vector fluid velocity. Each cell donates the portion of the fluid variable which overlaps its downstream neighbor (see Yanagita & Kaneko 1995 for a first-order implementation of this method). This method avoids the violations in certain multi-dimensional conservation rules known to plague schemes using the operator splitting technique developed by Strang (1968).

We adopt a planar approximating function for the fluid variable in each cell:

$$A^{i,j}(x,y) = \overline{A}^{i,j} + (x - x_0)a_x^{i,j} + (y - y_0)a_y^{i,j}. \tag{A1}$$

where $x_0$ and $y_0$ are the coordinates of the cell's center and $a_x^{i,j}$ and $a_x^{i,j}$ are the x and y slopes. In regions of the flow where the velocities are oriented in the same direction, a cell generally receives flux from three upstream neighbors. The amount of donated fluid is given by the double integral of the approximating function over the 2-D area of overlap between cells. For positive velocity components, the update rule becomes

$$\begin{aligned}
A^{i,j,t+1} = \\
& (1 - v_x^{i,j})(1 - v_y^{i,j})[A^{i,j} - \tfrac{1}{2}(v_x^{i,j}a_x^{i,j} + v_y^{i,j}a_y^{i,j})] + \\
& v_x^{i-1,j}(1 - v_y^{i-1,j})[A^{i-1,j} + \tfrac{1}{2}((1 - v_x^{i-1,j})a_x^{i-1,j} - \\
& \quad v_y^{i-1,j}a_y^{i-1,j})] + \\
& v_y^{i,j-1}(1 - v_x^{i,j-1})[A^{i,j-1} + \tfrac{1}{2}((1 - v_y^{i,j-1})a_y^{i,j-1} - \\
& \quad v_x^{i,j-1}a_x^{i,j-1})] + \\
& v_x^{i-1,j-1}v_y^{i-1,j-1}[A^{i-1,j-1} + \\
& \quad \tfrac{1}{2}((1 - v_y^{i-1,j-1})a_y^{i-1,j-1} - \\
& \quad (1 - v_x^{i-1,j-1})a_x^{i-1,j-1})]. 
\end{aligned} \tag{A2}$$

Similar expressions result for velocity components with negative or mixed signs.

When velocities oppose one another a modification must be made to prevent flow crossing. We introduce this modification by means of an example. Assume that $v_i > 0$ and $v_{i+1} < 0$. For this example we define the gas mass which is to be donated from cell $i$ to cell $i+1$ to be $m_i$. Similarly, $m_{i+1}$ is the mass donated from $i+1$ to $i$. We stipulate that the fluid in cell $i$ is successfully transferred to $i+1$ only if it has the greater momentum, i.e. $m_i|v_i| > m_{i+1}|v_{i+1}|$; otherwise the opposing fluid parcel sweeps up the former and the mass and momentum of each are donated to the local cell. Thus, opposing flows are prevented from streaming past each other and mass and momentum are conserved.

The slopes of the approximating functions are determined by first order differencing (van Leer 1977)

$$\begin{aligned}
a_x^{i,j} &= \tfrac{1}{2}(A^{i,j+1} - A^{i,j-1}) \\
a_y^{i,j} &= \tfrac{1}{2}(A^{i+1,j} - A^{i-1,j}).
\end{aligned} \tag{A3}$$

In regions near strong discontinuities, the slopes given by this method can cause the approximating function to overshoot, leading to negative densities and/or nonphysical oscillations. A monotonicity condition due to van Leer (1977) is imposed to limit the slope of the linear function, preventing it from exceeding the range of densities of the neighboring cells. In addition, if the cell is an extremum of the surrounding neighbors, the slope is set to zero. This condition is applied to the x- and y-axes independently, ensuring that the *average* value of the approximating plane along the cell edges is bounded by the neighbors.

$$a^i = \begin{cases} \min[a^i, 2|\overline{A}^{i+1} - \overline{A}^i|, 2|\overline{A}^i - \overline{A}^{i-1}|] \\ \qquad \text{if } \operatorname{sgn}[a^i] = \operatorname{sgn}[\overline{A}^i - \overline{A}^{i-1}] \\ 0 \qquad \text{otherwise} \end{cases} \tag{A4}$$

Near strong discontinuities, the monotonicity condition substantially reduces the approximating slopes. A further constraint is placed on the slopes to prevent any portion of the approximating plane from extending above or below the neighboring cells.

The stability of this scheme is discussed in van Leer (1977) and yields the standard Courant-Friedrichs-Levy (CFL) condition on the velocity $|v| \le 1$.

By expanding the dispersion equation to fourth order in $k\Delta x$ and second order in $\omega \Delta t$ we find the following algebraic dispersion relation

$$\omega \approx \frac{(k\Delta x)^4}{8 \Delta t}[(1 - v^3) - 2v(1 - v)]v. \tag{A5}$$

The steep $k^4$ dependence is the signature of the operator $\nu_4 \partial^4/\partial x^4$, where the effective diffusion coefficient $\nu_4$ takes the form

$$\nu_4 = -\frac{1}{12}[3(1 - v^3) - 6v(1 - v)]v. \tag{A6}$$

Near discontinuities when the monotonicity condition limits the slopes $a^i$, numerical diffusion is increased. In the limiting case where the slopes of the approximating functions are identically zero, the numerical diffusion is of the form $\nu_2 \partial^2/\partial x^2$, where the effective diffusion coefficient is

$$\nu_2 = \frac{1}{2}|v|(1 - |v|). \tag{A7}$$

In each case, the numerical diffusion coefficient depends on the fluid velocity, reaching a maximum at $v = 1/2$ and going to zero at $v = 0, 1$.

We have compared the evolution of a square wave for the first- and second-order schemes. We find both from an analysis of the simulations and a linearized analysis that the action of the second-order diffusion operator causes the length scale to evolve as $l \propto t^{1/4}$ compared to the $l \propto t^{1/2}$ scaling for the $\partial^2/\partial x^2$ operator.

Since the model equations do not include pressure, standard shock tube solutions cannot be used to test the code. Instead, we used the analytic solution of the Burgers equation for a rarefaction shock. Details are given in Chappell (1997). We find that density spikes are bounded by two accretion shocks. Between the shocks the average gas velocity is roughly constant and equal to the the velocity of the shocks. The emergence of double shocks arises from diffusion in the continuity equation. This can be seen by including diffusion terms in Eq. 1 and Eq. 2 and neglecting the star formation forcing term. The equations become

$$\frac{\partial \rho}{\partial t} + \nabla \cdot (\rho \vec{v}) = \nu_2 \nabla^2 \rho \tag{A8}$$

$$\frac{\partial \rho \vec{v}}{\partial t} + \nabla \cdot (\vec{v} \rho \vec{v}) = \nu_2 \nabla^2 \rho \vec{v} \tag{A9}$$

where we adopt the second-order derivatives in the numerical diffusion term since we are interested in the behavior around the shocks. Substituting Eq. A8 into Eq. A9, we have

$$\frac{\partial \vec{v}}{\partial t} + \vec{v} \cdot \nabla \vec{v} = \nu_2 \nabla^2 \vec{v} + 2(\nabla \log \rho)(\nabla \cdot \vec{v}) \quad \text{(A10)}$$

The first term on the right hand side represents a numerical viscosity. This term and the Lagrangian derivative on the left hand side of the equation constitute the Burgers equation. This term acts to smooth out the shock front. The second term on the right hand side produces the double shock. At the leading edge of the density enhancement, both the gradients of the density field and velocity field are negative so their product is positive and the velocity field is locally raised. At the trailing edge, the gradients in the density and velocity fields are positive and negative respectively, so their product is negative. This results in lowering the velocity field. Together these two effects lead to the double shock structure in what would be a single shock in the Burgers equation. We do not expect this effect to substantially influence the dynamics since the width of the density enhancement is found to grow only as $\lambda \propto t^{1/4}$ according to the fourth-order diffusion operator. Thus, the double-shock structure should remain as a relatively local phenomenon.

Myhill & Boss (1993) have pointed out that this scheme is actually only first order accurate when the velocity field varies in space or time. They derive correction terms to make the scheme truly second order accurate. These terms effectively allow the cells to be stretched depending on the space and time derivatives of the local velocity field. Since we are interested in studying the qualitative behavior of our systems, we do not adopt these more computationally expensive correction terms. Indeed, Yanagita & Kaneko (1995) have achieved remarkable success in using a truly first-order donor-cell scheme to study the scaling behavior and pattern formation in Rayleigh-Bernard convection.

## APPENDIX B: NUMERICAL DIFFUSION

To a first approximation the diffusion along the axes is separable and we may write the numerical diffusion coefficients near discontinuities as

$$\begin{aligned}
D_{2\,x} &= \tfrac{1}{2}|v_x|(1-|v_x|) \\
D_{2\,y} &= \tfrac{1}{2}|v_y|(1-|v_y|)
\end{aligned} \quad \text{(B1)}$$

and in smooth regions as

$$\begin{aligned}
D_{4\,x} &= \tfrac{1}{12}|v_x|[(1-|v_x|)^3 - 2|v_x|(1-|v_x|)] \\
D_{4\,y} &= \tfrac{1}{12}|v_y|[(1-|v_y|)^3 - 2|v_y|(1-|v_y|)].
\end{aligned} \quad \text{(B2)}$$

The velocity dependences of these coefficients are similar, peaking at $v = 1/2$ and going to zero at $v = 0, 1$.

This diffusion can be highly anisotropic. For example, a density spike moving along the x axis with $v_x = 1/2$ diffuses along the x axis but not in the y direction. Thus, disturbances moving along either axis will be stretched along their direction of motion. On the other hand, a particle moving along the diagonals with $|v_x| = |v_y|$ will retain its shape since the numerical diffusion coefficients are equal.

We have found that by adding diffusion locally only along the direction with the lesser numerical diffusion coefficient, the anisotropy can be effectively reduced without substantially increasing the overall diffusivity of the integration scheme. The amount each cell diffuses into the neighboring cells is proportional to the value of the approximating function at the cell walls given by $A^i_{\text{left}} = A^i - a^i/2$ and $A^i_{\text{right}} = A^i + a^i/2$. Thus, the update rule for the cross diffusion operator may be written as

$$A^{i,t+1} = A^{i-1}(1-2\epsilon) + \epsilon(A^{i-1} - \tfrac{1}{2}a^{i-1}) + \epsilon(A^{i+1} - \tfrac{1}{2}a^{i+1}) \quad \text{(B3)}$$

where $\epsilon$ is related to the diffusion coefficient. This expression may be rewritten as

$$A^{i,t+1} = A^i + \epsilon(A^{i+1} - 2A^i + A^{i-1}) - \tfrac{1}{2}\epsilon(a^{i+1} - a^{i-1}). \quad \text{(B4)}$$

When the monotonicity condition limits the slopes to near zero, the last term becomes small and the diffusion takes the form of the standard finite difference approximation of the $\partial^2/\partial x^2$ operator (the second term on the right hand side of Eq. B4). When the monotonicity condition has no effect, the cross diffusion takes the form

$$A^{i,t+1} = A^i + \tfrac{1}{4}\epsilon(-A^{i-2} + 4A^{i-1} - 6A^i + 4A^{i+1} - A^{i+2}) \quad \text{(B5)}$$

where the approximating slopes (Eq. A3) have been substituted in. This is a standard finite-difference representation of the $\partial^4/\partial x^4$ operator. Applying Eq. B5 directly to rugged fluid landscapes can lead to negative densities and instabilities; however Eq. B4 subject to the monotonicity condition ensures a stable scheme and positive definite densities for $\epsilon < 1/2$.

The resulting effective numerical diffusion coefficient along either axis may be written as

$$D_{\text{eff}} = \tfrac{1}{4}|v_r|[(1-|v_r|)^3 - 2|v_r|(1-|v_r|)] \quad \text{(B6)}$$

where

$$v_r = \begin{cases} |v| & \text{if } |v| < 1/2 \\ 1/2 & \text{otherwise} \end{cases} \quad \text{(B7)}$$

Condition B7 ensures that the diffusion coefficient is isotropic, depending only on the velocity amplitude. The resulting numerical diffusion increases with velocity for $|v| < 1/2$ and remains at a constant value for larger velocities. The transverse diffusion which must be added to the intrinsic numerical diffusion of the donor-cell scheme so that the effective diffusion is independent of direction is given by

$$\begin{aligned}
D_x^{\text{trans}} &= D_{\text{eff}} - D_{4\,x} \\
D_y^{\text{trans}} &= D_{\text{eff}} - D_{4\,y}
\end{aligned} \quad \text{(B8)}$$

where $D_{4\,x}$ and $D_{4\,y}$ are the numerical diffusion coefficients given in Eq. 22.

The transverse diffusion terms (Eq. B8) were implemented to reduce the anisotropic numerical diffusion effects inherent to multidimensional donor-cell advection schemes of the form given in equation A3. We devised two tests to evaluate how well these terms preserve the shape of moving density disturbances. In the first test, a square density enhancement was advected across the lattice at a given speed for a number of propagation directions. Using the unmodified advection scheme, the disturbances were often severely stretched along the x or y axis. We found that the inclusion of the transverse diffusion term preserved the shape of the original disturbance for all velocities examined. In the second test, the gas density along a circular, radially expanding ring was examined. In the unmodified advection scheme, the ring rapidly fragmented along the grid axes, while the inclusion of the transverse diffusion terms produced a substantially smoother density profile along the ring. For a detailed description of these and other tests see Chappell (1997).

## APPENDIX C: GRAVITATIONAL STABILITY OF AN EXPANDING, ACCRETING SHEET

We consider the gravitational stability of a two-dimensional sheet which is accreting and expanding. This analysis follows the work of Elmegreen (1994) and Comeron & Torra (1994) who were interested in the stability of a thin supernova remnant or wind-blown shell expanding into a uniform medium. Since the present analysis slightly extends their work to allow independent expansion and accretion rates, we include some of the steps in obtaining our results. It is necessary to derive the more general form because the "environmental" density and velocity fields in our simulations quickly distort the shells, preventing the application of Elmegreen's instability condition which requires knowledge of the radius and age of the shell and which assumes a homogeneous ambient gas distribution. In this derivation, instabilities related to bending modes of the shell such as those investigated by Vishniac (1983) are not considered. As suggested by Elmegreen (1994), such modes may affect the internal velocity dispersion of the shell.

The equations governing flows in the sheet of surface density $\mu$ are

$$\frac{\partial \mu}{\partial t} = -\nabla \cdot (\mu \mathbf{v}) + A \quad (C1)$$

$$\mu \frac{\partial \mathbf{v}}{\partial t} = -\nabla P - \mu \mathbf{v} \cdot \nabla \mathbf{v} + \mu \nabla \Phi - \mathbf{v} A \quad (C2)$$

$$\nabla \cdot \Phi = -4\pi G \rho. \quad (C3)$$

where $P$ is the gas pressure, $\Phi$ is the gravitational potential and $A$ is the accretion rate onto the sheet. In addition, we assume an isothermal equation of state.

We assume that the unperturbed surface density $\mu_0$ and pressure $P_0$ are uniform and that the sheet is expanding so that the divergence of the unperturbed velocity field is nonzero, i.e. $\nabla \cdot \mathbf{v_0} \neq 0$. Locally, however, we set $\mathbf{v_0} = 0$. The unperturbed surface density satisfies

$$\frac{\partial \mu_0}{\partial t} = A - \mu_0 \nabla \cdot \mathbf{v_0}. \quad (C4)$$

The first term on the right hand side is the rate of accretion of new material while the second term is the rate at which the local surface density decreases due to expansion of the sheet. The perturbed variables satisfy the following continuity and momentum equations:

$$\frac{\partial \mu_1}{\partial t} = -\mu_0 \nabla \cdot \mathbf{v_1} - \mu_1 \nabla \cdot \mathbf{v_0} \quad (C5)$$

$$\mu_0 \frac{\partial \mathbf{v_1}}{\partial t} = -c^2 \nabla \mu_1 - \mu_0 \mathbf{v_1} \nabla \cdot \mathbf{v_0} - 2\pi i G \mu_1 \mu_0 - \mathbf{v_1} A \quad (C6)$$

where we have assumed an isothermal equation of state and substituted $\nabla \Phi_1 = -2\pi i G \mu_1$.

Assuming the perturbations are of the form $e^{\omega t - ikx}$, the dispersion relation is found to be

$$(\omega + \nabla \cdot \mathbf{v_0} + A/\mu_0)(\omega + \nabla \cdot \mathbf{v_0}) = -c^2 k^2 + 2\pi G \mu_0 k \quad (C7)$$

In the limit that accretion and stretching go to zero, the dispersion relation for the infinite sheet is recovered. The action of accretion and stretching do not affect the wavelength of the fastest growing mode since these terms do not contain a $k$ dependence; they just retard the growth rate. Substituting the wave number of the fastest growing mode into the dispersion relation, we find that the condition for significant collapse (meaning that the growth rate exceeds a prescribed value $\omega_c$) becomes

$$\left(\frac{\pi G \mu}{c}\right)^2 > (\omega_c + \nabla \cdot \mathbf{v_0} + A/\mu_0)(\omega_c + \nabla \cdot \mathbf{v_0}). \quad (C8)$$

This dispersion relation reduces to the relations found by Elmegreen (1994) and Comeron & Torra (1994) to within proportionality constants for the simpler case of a spherical shell expanding into a uniform medium. The effect of shell expansion or "stretching" ($\nabla \cdot \vec{v}_0 > 0$) is to make collapse more difficult, as expected.